\documentclass[a4paper,12pt]{article}
\usepackage{mDraft}

\renewcommand{\eqref}[1]{Eq.~(\textcolor{red}{\ref{#1}}) }

\def\ie{{\it i.e.}}

\newcommand{\bew}{\begin{widetext}}
\newcommand{\enw}{\end{widetext}}
\newcommand{\bee}{\begin{equation}}
\newcommand{\ene}{\end{equation}}
\newcommand{\bea}{\begin{eqnarray}}
\newcommand{\ena}{\end{eqnarray}}
\newcommand{\bes}{\begin{subequations}}
\newcommand{\ens}{\end{subequations}}

\def\to{\rightarrow}

\def\met{E_T \hspace*{-1.1em}/\hspace*{0.5em}}

\def\call{\mathcal{L}}
\def\calm{\mathcal{M}}

\def\calo{\mathcal{O}}

\def\mev{\,{\rm MeV}}
\def\gev{\,{\rm GeV}}
\def\tev{\,{\rm TeV}}

\def\fb{\,{\rm fb}}

\def\ab{\,{\rm ab}}

\def\iab{$\,{\rm ab}^{-1}$}

\title{
Exploring Four Fermion Contact Couplings of a Dark Fermion and an Electron at Hadron Colliders and Direct Detection Experiments
}
\author[a$\dag$]{Kai Ma}
\email[$\dag$]{kai@xauat.edu.cn}
\affiliation[a]{Faculty of Science, Xi'an University of Architecture and Technology, Xi'an, 710055, China}

\abstract{\noindent
Both collider searches and direct detections are promising approaches 
to probe fermionic dark matter. In this paper, 
we study signatures of four-fermion contact operators involving a dark fermion, 
an electron, and a quark pair. We show that 
the mono-electron production channel at hadron colliders 
can provide strong constraints. Associated productions of a charged electron 
with a photon/jet and missing energy are also studied. 
Using current LHC data at $\sqrt{s} = 13 \,\text{TeV}$, 
the lower bound on the energy scale of the (axial-)vector operator 
can reach 12 TeV for a massless dark fermion. 
This can be further improved to about 24 TeV at the HE-LHC 
with $\sqrt{s} = 25 \,\text{TeV}$ and a total luminosity of $20 \, \text{ab}^{-1}$.
For direct detections, the signal operators can induce $\beta^\pm$ decays. 
For the induced $\beta^-$ decay, we find that the constraints are weaker than 
those from collider searches in almost all of the parameter space, 
and the accessible parameter space has already been excluded 
by current LHC data. In the case of a relatively heavy dark fermion (a few MeV), 
the induced $\beta^+$ decay is more sensitive than collider searches. 
Despite the advantage of collider searches that a much wider range of dark fermion masses can be investigated, they can also provide complementarity to direct detections.
}

\begin{document}
\maketitle

\section{Introduction}
\label{sec:intro}
The astrophysical and cosmological evidence for Dark Matter (DM) \cite{Bertone:2004pz,Young:2016ala,Arbey:2021gdg} has motivated numerous experimental searches for its physical properties beyond gravitational effects. Interactions between DM and Standard Model (SM) particles can be probed indirectly by observing signals of DM annihilations into SM particles in high-density regions of the universe \cite{Leane:2020liq,Slatyer:2021qgc}. For instance, significant signals of anomalous lepton and photon flux can arise from DM annihilation \cite{Bouquet:1989sr,Baltz:2002we,John:2021ugy,Bi:2009uj,Ibarra:2009bm}. However, no signal has been observed so far beyond its gravitational interactions \cite{Roszkowski:2017nbc,deDiosZornoza:2021rgw,Arguelles:2023nlh}.

Alternatively, DM can be probed via scattering off nuclei \cite{Goodman:1984dc,Batell:2014yra,Dror:2019onn,Dror:2019dib}, electrons \cite{Dror:2020czw}, or phonons \cite{Mitridate:2023izi} of atomic targets in terrestrial experiments, also known as direct detections \cite{Liu:2017drf,Billard:2021uyg,Buchmueller:2014yoa}. Scattering processes can be either elastic or inelastic. Elastic scattering off nuclei has provided strong constraints on interactions between nucleons and DM with masses in the GeV-TeV range \cite{LZ:2022lsv}. If the DM has multiple particle components, mass gaps between these particles can be converted into kinematic energy of the final states. Such inelastic scattering can probe both heavy and light DM particles, provided the mass gap is sufficiently large. Nevertheless, inelastic scattering can also occur for a single DM particle via a process called absorption \cite{Dror:2019onn,Dror:2019dib,Dror:2020czw}, in which the DM is converted into a SM neutrino by the target, and its mass is completely transferred to the final states. Hence, this process is naturally sensitive to light DM.

On the other hand, colliders can also probe DM \cite{Boveia:2018yeb,Gori:2022vri,Lagouri:2022ier,Penning:2017tmb,Buchmueller:2014yoa}. Since DM couples very weakly to known particles, it appears as missing energy in collider detectors. Collider searches therefore focus on signatures with large missing energy. The advantage of this approach is its ability to probe a much wider range of DM masses, limited only by the collider's center-of-mass energy. This makes it particularly promising for heavy dark particles predicted by UV-complete models of the Dark Sector \cite{delAguila:2014soa,Buckley:2015cia,Farzan:2010mr,Su:2009fz,Kahlhoefer:2017dnp,Bartels:2012ex,Kundu:2021cmo,Barman:2021hhg,Liang:2021kgw,Freitas:2014jla,
Dreiner:2012xm,Habermehl:2020njb,Kalinowski:2021tyr,Bharadwaj:2020aal}. 

Although direct detections and collider searches represent completely different mechanisms for hunting DM particles, they are strongly correlated in an inevitable way. The hadron-level operators that induce DM scattering off atomic targets must be matched to parton-level interactions responsible for its direct production at colliders \cite{Boveia:2018yeb,Gori:2022vri,Lagouri:2022ier,Penning:2017tmb}. Therefore, a combined analysis of these two approaches is essential to provide complementary constraints on the model parameters \cite{Ge:2023wye,Belyaev:2018pqr}.

If DM stability is preserved by a discrete symmetry such as \(\mathbb{Z}_2\), elastic scattering with a small mass and inelastic scattering with a much larger mass can be significant in direct detection experiments, while at colliders DM particles are produced in pairs. In contrast, if DM is always accompanied by a neutrino in interactions with other known particles, the most sensitive channel in direct detections is the absorption process \cite{Goodman:1984dc,Batell:2014yra,Dror:2019onn,Dror:2019dib,Dror:2020czw}, and collider missing energy consists of DM and a neutrino. Even though they share the same final-state topology \cite{Dreiner:2013vla,Belwal:2017nkw,Dror:2019dib}, the kinematic properties of the missing energy can differ significantly \cite{Ge:2023wye}. As a result, constraints on the model parameters—particularly those from combined direct detection and collider searches—can vary substantially. Furthermore, DM can remain stable with the correct relic abundance if it is light enough that its decay width is sufficiently small \cite{Ge:2022ius,Ge:2023wye}.

The couplings between femionic DM, neutrinos, and matter particles have been extensively studied. Here, we focus on the case where the dark fermion is always accompanied by a charged lepton. These two distinct interactions yield distinctive signals in both direct detection and collider search experiments. For instance, DM-lepton couplings can mimic induced beta decay in direct detections and produce mono-lepton events or associated productions of a lepton with an energetic photon or jet at hadron colliders.

In the effective field theory (EFT) framework, the lowest-order interactions with matter particles are described by four-fermion contact operators. Via Fierz transformation, four-fermion contact couplings involving a DM-neutrino and a lepton pair are equivalent to those involving a DM-lepton and a lepton-neutrino. Hence, this paper focuses on the phenomenology of four-fermion contact couplings between a DM-lepton pair and a quark pair. We will investigate these interactions in both direct detection and collider experiments, as well as their complementary constraints.

The rest of this paper is organized as follows. In Sec.~\ref{sec:eft}, we discuss the essential properties of the charged four-fermion contact interactions. In Sec.~\ref{sec:Cons:LHC13}, we study signal properties of the mono-lepton process and associated production of a lepton with a photon or a jet, as well as constraints from current LHC data. Details of the mono-\(\ell\), \(\ell + \gamma\), and \(\ell\)+jet productions are given in Subsections \ref{sec:Monol}, \ref{sec:LepPho}, and \ref{sec:LepJet}, respectively. Sensitivities achievable at HE-LHC are studied in Subsection \ref{sec:Future}. Constraints from direct detections in induced $\beta$-decay are studied in Sec.~\ref{sec:InvBetaDecay}. Our conclusions are given in Sec.~\ref{sec:conclusion}.

\section{Effective Operators}
\label{sec:eft}
For a fermionic dark particle, 
the effective interaction operator involving matter particles 
can be described at the lowest order by four-fermion contact operators  
\cite{Cepedello:2023yao,Belwal:2017nkw,Dreiner:2013vla,Falkowski:2015krw}. 
If the couplings between dark fermion and quark pairs are preserved by some symmetry, 
signals can be probed via either elastic scattering of dark fermions off atomic nuclei \cite{Goodman:1984dc} or mono-X processes (\(X = \gamma\), \(Z/W\) bosons, jets, etc.) at hadron colliders  \cite{Bernreuther:2018nat,Krovi:2018fdr,Liew:2016oon,Bhattacharya:2022qck}. 
On the other hand, the dark fermion can participate via mixing with neutrinos  \cite{Dror:2019onn,Dror:2019dib}. The neutral current formed by dark fermion-neutrino pairs can induce inelastic scattering via dark fermion absorption by nuclear targets ~\cite{Dror:2019onn,Dror:2019dib,Li:2022kca}.
In addition to the neutral current, the dark fermion can couple to quark pairs via charged currents formed with charged leptons \cite{Dror:2019onn,Dror:2019dib}. 
These couplings give rise to nontrivial signals --- for example, 
charged currents can make significant contributions to inverse $\beta$-decay \cite{Dror:2019onn,Dror:2019dib}. 
Here, we focus on four-fermion contact operators composed of a dark fermion, 
a charged lepton, and a quark pair.

The general four fermion contact interaction operators can be found in Ref.~\cite{Lee:1956qn}.
Here we focus on a minimal scenario about the dark fermion, 
where either left- or right-handed dark fermions couple to the charged lepton. 
In consideration of that the possible mixing effect between the dark fermion 
and the neutrino can be easily taken into account for UV completed model building 
\cite{Dror:2019dib,Dror:2020czw,Li:2022kca},
we simply assume that the charged lepton 
(which is in the same isospin doublet with the neutrino) is left handed. 
Within this assumption, the EFT interaction Lagrangian can be explicitly written as,
\begin{equation}
\label{eq:effo}
\begin{aligned} 
- \call_{S} =
\frac{1}{\varLambda_{S}^2} \mathcal{O}_{S} & \equiv
\frac{1}{\varLambda_{S}^2}
\Big[\overline{d} u \Big] \Big[\overline{\chi} \gamma_{L} \ell \Big] \,,
\\ 
- \call_{P} =
\frac{1}{\varLambda_{P}^2} \mathcal{O}_{P} 
& \equiv
 \frac{1}{\varLambda_{P}^2}
\Big[\overline{d} (i \gamma_{5} ) u \Big] 
 \Big[\overline{\chi} \gamma_{L} \ell \Big]  \,,
\\ 
- \call_{V} =
\frac{1}{\varLambda_{V}^2} \mathcal{O}_{V} 
& \equiv
\frac{1}{\varLambda_{V}^2}
\Big[\overline{d} \gamma_{\mu} u \Big] 
\Big[\overline{\chi} \gamma^\mu \gamma_{L} \ell \Big] \,,
\\ 
- \call_{A} =
\frac{1}{\varLambda_{A}^2} \mathcal{O}_{A} 
& \equiv
 \frac{1}{\varLambda_{A}^2}
\Big[\overline{d} \gamma_{\mu}\gamma_5 u \Big] 
\Big[\overline{\chi} \gamma^\mu \gamma_{L} \ell \Big] \,,
\\ 
- \call_{T} =
\frac{1}{\varLambda_{T}^2}  \mathcal{O}_{T}
& \equiv
\frac{1}{\varLambda_{T}^2}
\Big[\overline{d} \sigma_{\mu \nu} u \Big] 
\Big[\overline{\chi} \sigma^{\mu \nu} \gamma_L \ell \Big] \,,
\end{aligned}
\end{equation}
where \(\Lambda_i\) are independent energy scales characterizing possible fundamental new physics related to the corresponding operators \(\mathcal{O}_i\). 
One may note that both left- and right-handed dark fermions appear in the above operators. However, the right-handed dark fermion \(\chi_R\) can be written as the left-handed state of its charge-conjugate state, \(\chi^c_L\). Hence, the right-handed dark fermion \(\chi_R\) in the first, second, and fifth equations of \eqref{eq:effo} can be replaced by \(\chi^c_L\). Since the dark fermion is neutral, this replacement has no effect. Hence, only left-handed dark fermions take part in the effective interactions. The same applies to the left-handed dark fermion \(\chi_L\) and its charge-conjugate state, \(\chi^c_R\). In this sense, a minimal scenario about the dark fermion is studied in this work.It is worth pointing out that different assumptions of the lepton chirality can potentially have non-trivial effects on the differential distributions of the scattering events. In our case, since both leptons and anti-leptons in the final-state configurations of the processes studied in this work are taken into account, and the dark fermion is measured inclusively, the differential distributions to be used for calculating the significances do not change. For direct detection experiments, chirality-induced phase differences after non-relativistic approximation are also unobservable.
In this sense, we study a minimal scenario for the dark fermion where only left-handed charged leptons are involved.
The above parameterization is complete in the sense that all five independent Lorentz structures of the quark bilinear are taken into account.

Due to the running of the EFT operators defined in \eqref{eq:effo}, 
operators with different Lorentz structures at high energy scales 
can mix into a combination at low energy  
\cite{Hill:2011be,Frandsen:2012db,Vecchi:2013iza,Crivellin:2014qxa,DEramo:2014nmf,DEramo:2016gos,Bishara:2017pfq,Belyaev:2018pqr}. 
The mixing effect becomes numerically significant for large couplings 
between the dark fermion and heavy quarks (particularly the top quark)  \cite{Hill:2011be,Frandsen:2012db,Vecchi:2013iza,Crivellin:2014qxa,DEramo:2014nmf,DEramo:2016gos,Bishara:2017pfq,Belyaev:2018pqr}. 
Here, we have assumed that the above effective operators apply only to first-generation quarks. 
Within this approximation, the mixing effect can be safely neglected, 
and we will study interactions with third-generation quarks in separate work.
Furthermore, the dark fermion can in principle couple to all three charged leptons, 
with couplings that are either universal or independent. 
However, including more leptons leads to diverse phenomenologies 
in direct detection and collider experiments, as well as dark fermion decay, 
which is crucial for reproducing the correct relic abundance. 
For simplicity, we assume the charged lepton is the electron. 
These assumptions are necessary 
to ensure a consistent combination of collider and direct detection constraints, 
where only light quarks and the electron are relevant.

Furthermore, signal topologies of the four-fermion operators at colliders 
depend on the invisibility or decay width \(\Gamma_{\chi}\) of the dark fermion. 
If the dark fermion decays quickly, it can be visible in the detector. 
In this paper, we focus on signatures of an invisible dark fermion. 
Hence, a study of dark fermion stability is necessary. Within our assumption, 
the dark fermion can decay only through the following channel:
\bee
\chi \to e^- + u + \bar{d} \,.
\ene
The total decay widths are given as,
\bea
\varGamma_{S/P} &=& \frac{ m_\chi^5 }{1024 \pi^3 \varLambda_{S/P}^4} \,,
\\[2mm]
\varGamma_{V/A} &=& \frac{ m_\chi^5 }{ 256 \pi^3 \varLambda_{V/A}^4} \,,
\\[2mm]
\varGamma_{T} &=& \frac{ 3 m_\chi^5 }{ 128 \pi^3 \varLambda_{T}^4} \,.
\ena
One can see that the total decay widths scale with \(m_\chi^5\) for all operators. 
In the case that the dark fermion \(\chi\) is very light, 
e.g., \(m_\chi < \varLambda_{\rm QCD}\), 
the above decay channels are forbidden due to quark-pair condensation. 
Conversely, signals of four-fermion contact couplings can appear 
in charged pion decays, \(\pi^{\pm} \to e^{\pm}\chi\) \cite{PIENU:2017wbj}. 
Due to large backgrounds from muon decays, 
these searches are sensitive only to dark fermions with \(m_\chi > 60\ \text{MeV}\). 
However, this region is outside our focus for light dark fermions 
(for direct detections via induced $\beta$-decay studied in Sec. \ref{sec:InvBetaDecay}). 
On the other hand, at a collider with (parton-level) center-of-mass energy 
\(\sqrt{\hat{s}}\), the typical decay length of the dark fermion is given by,
\bee
L_{\chi} 
= \gamma_{\chi} \tau_{\chi} 
= \frac{ \sqrt{\hat{s}} }{2 m_\chi \varGamma_{\chi} }  
\Big( 1 + \frac{ m_\chi^2 }{ \sqrt{\hat{s}} } \Big)\,.
\ene
For a relatively light dark fermoin, one has 
$L_{\chi}  \approx  \sqrt{\hat{s}}/(2 m_\chi \varGamma_{\chi} ) $.
For reference, 
we conservatively require that the typical decay length exceeds 1\,m.

\begin{figure}[h] 
\centering
\includegraphics[width=0.58\textwidth]{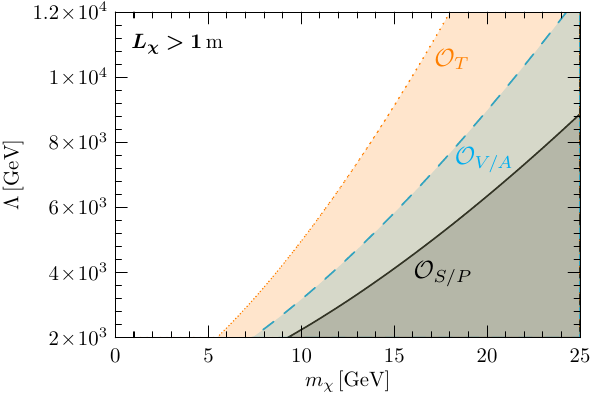}
\caption{\it 
Typical decay length of the dark fermion in the $m_\chi$-$\varLambda$ plane.
The parton-level center-of-mass energy is chosen as $\sqrt{\hat{s}}=1\tev$.
}
\label{fig:dcylen}
\end{figure}

Fig.~\ref{fig:dcylen} shows the regions with $L_\chi < 1\,$m for a typical 
parton-level center-of-mass energy at the LHC, $\sqrt{\hat{s}}=1\tev$.
One can see that for a dark fermion with mass \(m_\chi \lesssim 5\ \text{GeV}\), 
the assumption of invisibility is always valid. 
Such light dark fermions allow for a variety of thermal (and non-thermal) production mechanisms 
that can account for the observed dark matter relic abundance \cite{Dror:2019dib,Dror:2020czw}. 
In this paper, instead of discussing details of these production mechanisms, 
we study the signal properties at hadron colliders and direct detection experiments 
in a model-independent way.
For relatively heavier dark fermion, 
the invisibility constraint depends on the energy scale \(\Lambda_i\). 
However, we emphasize that this constraint is model-dependent 
rather than general, particularly when the dark fermion \(\chi\) 
is just one component of a broader dark sector  
\cite{Gori:2022vri,Marra:2019lyc,Deliyergiyev:2015oxa,Hofmann:2020wvr,Lagouri:2022ier}.

Given the superiority of high-energy colliders, \ie, 
their ability to search for much heavier particles, 
we will investigate signals of a dark fermion \(\chi\) across its full mass range, 
provided it is kinematically accessible. 
Only in Sec.~\ref{sec:InvBetaDecay}, 
where \(\chi\) is considered a potential DM candidate, 
is its mass less than \(10\ \text{MeV}\). Within this mass window, 
direct decay channels are forbidden by QCD color confinement. 
While off-shell decay mediated by \(\pi\)-mesons occurs, 
the decay width is strongly suppressed by electroweak couplings 
and off-shell kinematic factors:  
$\varGamma \sim G_{f}^2 f_\pi^2 (m_\chi/m_\pi)^4  m_\chi^5 /( 256 \pi^3 \varLambda^4 )$,
where \(G_{\text{F}}\) is the Fermi constant, \(f_\pi\) is the pion decay constant.
For a typical new physics energy scale \(\varLambda \sim 1\tev\) 
(and dark fermion with mass \(10\ \text{MeV}\)), 
the decay width is of the order of \(10^{-43}\gev\), 
corresponding to a lifetime of \(10^{17}\,\text{s}\). 
This is comparable to the age of the Universe, 
hence rendering the dark fermion effectively stable at higher energy scales.
On the other hand, decay channels can also emerge at the loop level; 
for instance, the four-fermion contact interactions investigated 
in Refs.~\cite{Dror:2019onn,Dror:2019dib,Dror:2020czw,Ge:2022ius} 
fall into this category. However, the dark fermion’s precise decay width 
hinges crucially on the UV realization of its neutrino-philic nature. 
A paradigmatic example is the axion (axion-like) portal model  
\cite{Fitzpatrick:2023xks,Dror:2023fyd,Anilkumar:2024tda}, 
which features non-trivial mixing between the dark fermion and neutrinos 
\cite{Dror:2019onn,Dror:2019dib,Dror:2020czw,Ge:2022ius,Li:2022kca,Ge:2023wye,Ma:2024gqj,Ge:2024euk}. 
Given this theoretical ambiguity, the task of systematically embedding 
these effective operators into a UV-complete framework 
lies outside the current work’s scope. 
We intend to address this in subsequent investigations.

\section{Constraints at the LHC}
\label{sec:Cons:LHC13}
At high-energy colliders, the above effective interactions may occur at energy scales exceeding the cutoff scale, potentially violating unitarity constraints as discussed in Refs.~\cite{Busoni:2013lha,Busoni:2014sya,Busoni:2014haa}. 
Under the assumptions that the momentum transfer \(|Q|\) remains 
below the mediator mass \(M_D\) (i.e., \(|Q| < M_D\)) 
and coupling strengths remain perturbative (\(< 4\pi\)), Refs.~\cite{Busoni:2013lha,Busoni:2014sya,Busoni:2014haa} show that 
events are reliable only for \(|Q| < 4\pi \Lambda\). 
Our analysis indicates that applying this momentum cutoff has a negligible impact 
on exclusion limits for two key reasons: 
(1) the \(4\pi \Lambda\) threshold is relatively weak, 
and (2) these events are naturally suppressed by proton parton distribution functions (PDFs), as detailed in Ref.~\cite{Ellis:2021dfa}. 
However, imposing a stricter cutoff (e.g., \(|Q| < \Lambda\)) 
could eliminate a significant fraction of events, 
thereby weakening exclusion limits, as noted in Ref.~\cite{Busoni:2014sya}.

On the other hand, it should be emphasized that 
without an explicit UV completion of the EFT framework, 
the cutoff scale cannot be rigorously determined. 
For model-independent studies, we advocate adopting the most conservative cutoff \(|Q| < 4\pi \Lambda\) consistent with perturbative validity. 
Moreover, the condition \(|Q| > m_{\rm DM}/(2\pi)\) is commonly adopted as a benchmark for EFT applicability. 
In the present study, the dark fermion \(\chi\) is treated as a viable DM candidate only when its mass falls in the \(\sim 1\)-\(100\ \text{MeV}\) range (see Sec.~\ref{sec:InvBetaDecay} for details), which inherently meets this criterion.
Therefore, the EFT description of collider searches remains valid provided that 
the cutoff scales \(\Lambda_i\) are higher than the parton-level center-of-mass energy \cite{Dreiner:2013vla}, 
or the mass of a possible mediator is significantly larger than the collision energy ~\cite{Busoni:2013lha,Busoni:2014sya,Busoni:2014haa}.

Signatures of charged-current operators differ entirely from those of neutral-current operators. 
For neutral-current operators, since the dark fermion and neutrino manifest as missing energy at colliders, 
the most sensitive probes are mono-$X$ production processes \cite{Bernreuther:2018nat,Krovi:2018fdr,Liew:2016oon,Bhattacharya:2022qck}, where $X$ is some visible particle. For instance, mono-photon  \cite{Abdallah:2015uba,daSilveira:2023hmt,Gabrielli:2014oya,Gershtein:2008bf,Hicyilmaz:2023tnr}, mono-$Z/W$  \cite{Abdallah:2015uba,Kawamura:2023drb,Yang:2017iqh,Abdallah:2019tpo,No:2015xqa,Bell:2012rg,Alves:2015dya,Wan:2018eaz,Bell:2015rdw}, and mono-jet processes \cite{Abdallah:2015uba,Claude:2022rho,Belyaev:2018ext,Bai:2015nfa}. In this case, total cross sections are proportional to \(\alpha_{EM}/\Lambda^4\) or \(\alpha_s/\Lambda^4\).
In contrast, for the charged-current operators defined in \eqref{eq:effo}, 
a visible charged lepton always appears in dark fermion production processes. 
As a result, the total cross section scales as \(1/\Lambda^4\), 
roughly two orders of magnitude larger. 
Hence, constraints on charged-current operators are much stronger than those on neutral-current ones. 
The most sensitive channel is clearly mono-lepton production. 
We also study details of associated productions of the dark fermion 
with a lepton and a photon or a jet. These processes can enhance constraints when combined and help distinguish signal interaction mechanisms. 
In addition, in our analysis, strong cuts on transverse momentum and missing energy are applied. In this kinematic region, both the total cross section and differential distributions are not heavily contaminated by parton showering effects. Thus, we did not consider these effects in our study.
Our numerical simulations are performed at parton level using the 
MadGraph \cite{Alwall:2014hca,Frederix:2018nkq} and 
FeynRules \cite{Alloul:2013bka} toolboxes in 
UFO format \cite{Degrande:2011ua}. 
We will show the validation details of our simulations below.

\subsection{Mono-Lepton Production}
\label{sec:Monol}
\begin{figure}[th]
\centering
\includegraphics[height=0.12\textheight]{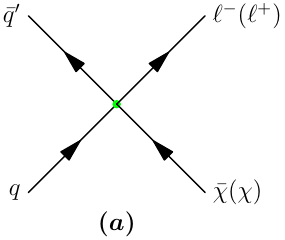}
\quad
\includegraphics[height=0.12\textheight]{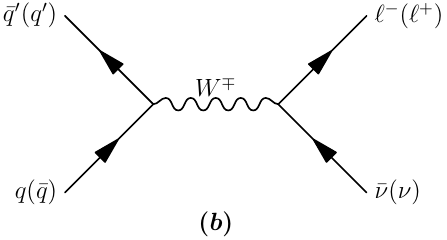}
\caption{\it 
Feynman diagrams of the mono-lepton process $pp \to \ell^\pm\slashed{E}_T + X$:
\textbf{(a)} is for the signal operators, 
\textbf{(b)} is for the irreducible background.
}
\label{fig:Feyn:Monol}
\end{figure}

Given our assumption that the dark fermion couples only to electrons, 
mono-lepton production corresponds specifically to mono-electron production 
at hadron colliders.,
\bee
p + p \;\to\; e^\pm + \slashed{E}_T + X\,,
\ene
where the transverse missing energy originates from the dark fermion 
\(\chi(\bar\chi)\) for the signal and from neutrinos for irreducible backgrounds. 
The corresponding parton-level Feynman diagrams are shown in Fig. \ref{fig:Feyn:Monol}: 
(a) for the signal and (b) for the irreducible background. 
Notably, mono-\(W^\pm\) production followed by leptonic decay is the dominant irreducible background. 
As a result, this background can be significantly suppressed by applying a kinematic cut on the transverse mass constructed from the electron and transverse missing energy. 
Hence, mono-electron production is expected to be the most sensitive channel 
for searching dark fermions at the LHC.

\begin{figure}[th]
\centering
\includegraphics[width=0.44\textwidth]{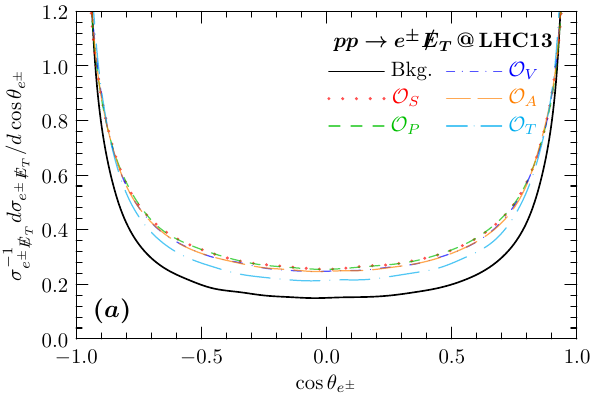}
\quad
\includegraphics[width=0.44\textwidth]{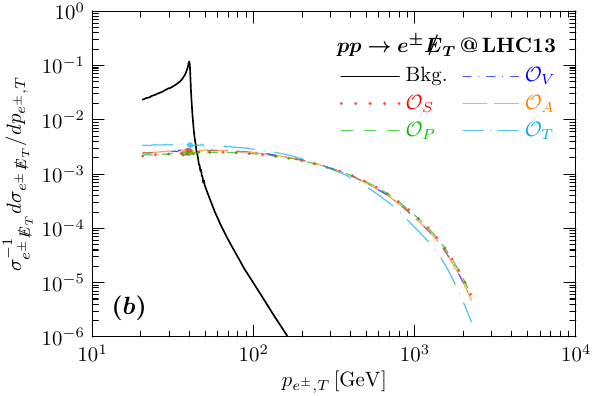}
\caption{\it 
Normalized distributions of the polar angle ($\theta_{e^\pm}$) \textbf{(a)} and 
transverse momentum ($p_{e^\pm, T}$) \textbf{(b)} of the outgoing electron
at generator level in the laboratory frame with center of mass energy $\sqrt{s} = 13\tev$. 
In the above panels, the signals (colorful non-solid curves) are shown for parameters 
$m_\chi = 0\gev$ and $\varLambda_i = 1\tev$, 
and the background (black-solid curve) stands for the irreducible
contribution from the channel $pp \to W^\pm \to e^\pm \nu$.
}
\label{fig:DXS:LX:LHC13}
\end{figure}

Figs. \ref{fig:DXS:LX:LHC13} (a) and (b) show normalized distributions 
of polar angle (\(\theta_{e^\pm}\)) and transverse momentum (\(p_{T,e^\pm}\)) 
for electrons in the laboratory frame 
at a center-of-mass energy of \(\sqrt{s} = 13\ \text{TeV}\), respectively. 
Signal event properties (colorful non-solid curves) are shown for parameters 
\(m_\chi = 0\ \text{GeV}\) and \(\Lambda_i = 1\ \text{TeV}\), 
while the irreducible background is depicted by the black solid curve.
From Fig.~\ref{fig:DXS:LX:LHC13} (a), one can see that 
regardless of the distinct Lorentz structures of the operators in \eqref{eq:effo}, 
electrons are dominantly produced in the forward and backward regions. 
This also holds for the irreducible background, 
as \(W^{\pm}\) bosons are primarily emitted along the incoming partons. 
Hence, differences in the polar angle distribution within the central region 
are negligible for both enhancing signal significance 
and distinguishing the Lorentz structures of signal operators.
In contrast, transverse momentum is a powerful observable 
for enhancing signal significance, as shown in Fig.~\ref{fig:DXS:LX:LHC13} (b). 
For irreducible background events, 
electrons emerge from \(W^{\pm}\) boson decays, 
so their transverse momentum is dominantly peaked 
at half the $W^{\pm}$ boson mass, {\it i.e.}, 
$p_{T,e^\pm} \sim m_W/2$. Above this threshold, 
the \(p_{T,e^\pm}\) distribution drops rapidly.
Signal electrons, however, are not constrained by such resonant decays. 
Their transverse momentum is limited only by the center-of-mass energy 
\(\sqrt{s}\) and PDF suppression at large energy fractions. 
As observed, the \(p_{T,e^\pm}\) distribution begins to decrease significantly 
only when \(p_{T,e^\pm}\) reaches hundreds of GeV.

The mono-lepton production channel has been searched 
by both the ATLAS \cite{ATLAS:2019lsy,ATLAS:2017jbq} 
and CMS \cite{CMS:2018hff} collaborations 
for searching heavy charged bosons. 
The observable used to discriminate between signal and background 
is the transverse mass of the charged lepton and missing transverse energy, defined as,
\bee
\label{eq:MT:LM}
m_T = \sqrt{ 2 p_{\ell, T} \slashed{E}_T (1 -  \cos\phi_{\ell\nu} ) } \,,
\ene
where \(\phi_{\ell\nu}\) is the azimuthal angle difference 
between the charged lepton and missing transverse momentum 
in the transverse plane. At parton level, 
the charged lepton and missing momentum are always back-to-back 
in the transverse plane, so the transverse mass simplifies to 
twice the charged lepton's transverse momentum, \ie, \(m_T = 2 p_{\ell, T}\).

\begin{figure}[th]
\centering
\includegraphics[width=0.48\textwidth]{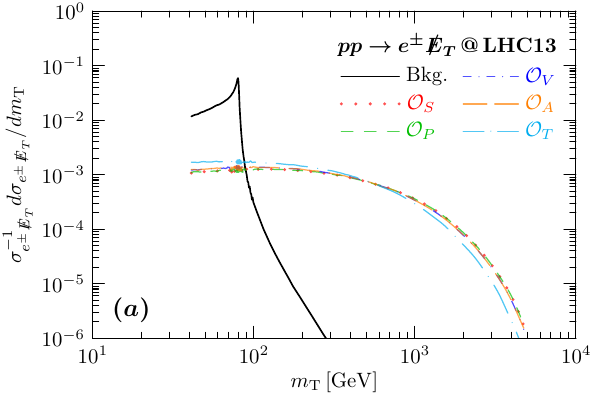}
\quad
\includegraphics[width=0.48\textwidth]{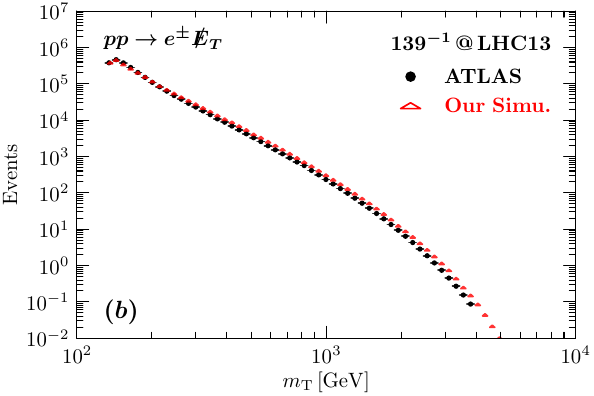}
\caption{\it 
\textbf{(a)}: 
Normalized distributions of the transverse mass ($m_T$) of the outgoing electron
and the missing transverse energy at generator level in the laboratory frame 
with center of mass energy $\sqrt{s} = 13\tev$. 
The signals (colorful non-solid curves) are shown for parameters 
$m_\chi = 0\gev$ and $\varLambda_i = 1\tev$, 
and the background (black-solid curve) stands for the irreducible
contribution from the channel $pp \to W^\pm \to e^\pm \nu$.
\textbf{(b)}: 
Validation of our simulation for the transverse mass ($m_T$) distribution 
of the irreducible background process $pp \to W^\pm \to e^\pm \nu$
at the LHC with center of mass energy $\sqrt{s} = 13\tev$ 
and a total luminosity $\call = 139\fb^{-1}$.
The experimental results (black dots) are taken from the Ref. \cite{ATLAS:2019lsy},
and our results (red triangles) have been renormalized by multiplying 
an overall constant such that the total number of events is matched.
}
\label{fig:LX:MT:LHC13}
\end{figure}
Fig. \ref{fig:LX:MT:LHC13} (a) shows normalized distributions of parton-level 
transverse mass at a center-of-mass energy of \(\sqrt{s} = 13\ \text{TeV}\) 
for signals (colorful non-solid curves) and the irreducible background 
(black solid curve). The irreducible background exhibits a clear peak at 
\(m_T = m_W\), with the production rate dropping sharply 
above this resonant peak, while signal distributions feature long tails. 
Here, we use the ATLAS's result --- with a luminosity of \(139\ \text{fb}^{-1}\) --- 
as a reference dataset to validate our simulations.

Events are selected with the requirements that 
both the missing transverse energy \(\slashed{E}_T\) 
and electron transverse momentum \(p_{T,e^\pm}\) exceed \(65\ \text{GeV}\), 
and electron rapidity satisfies \(|\eta_e| < 2.47\) 
and \(|\eta_e| \notin [1.37, 1.52]\) (to exclude the barrel-endcap transition region). 
The signal region is defined by \(m_T > 130\ \text{GeV}\).
Fig. \ref{fig:LX:MT:LHC13} (b) shows the validation of our simulation 
for the irreducible background process \(pp \to W^\pm \to e^\pm \nu\) 
in the signal region. For comparison, 
our result is renormalized by a universal scale factor 
to match the total number of irreducible background events, 
accounting for possible detector effects. 
The excellent agreement with ATLAS data demonstrates that 
the approximation of an overall normalization factor works well 
for both total event counts and differential distributions.

The expected exclusion limits at the LHC13 are estimated by calculating the following $\chi^2$,
\bee
\chi^2  =  \sum_{i} 
\left[  \frac{ \epsilon_D \cdot N^{\rm S}_{i} }{ \sigma^{\rm ATLAS}_{i} } \right]^2\,,
\ene
where \(\sigma_i^{\rm ATLAS}\) is the experimental uncertainty 
in the number of events in the $i$-th bin of the transverse mass \(m_T\) 
(from the ATLAS paper \cite{ATLAS:2019lsy}), 
\(N_i^{\rm S}\) is the number of signal events in the $i$-th bin, 
and \(\epsilon_D\) is the detector efficiency, as previously explained.

\begin{figure}[h]
\centering
\includegraphics[width=0.48\textwidth]{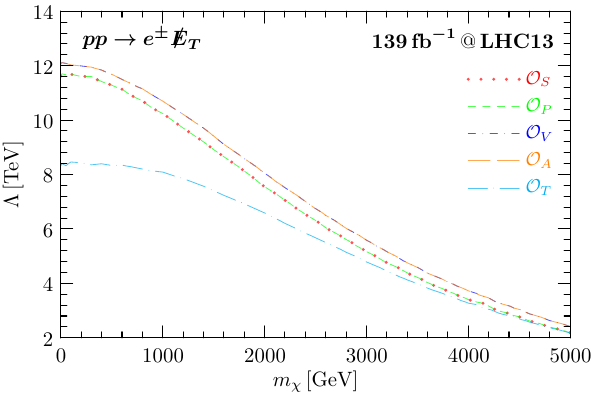}
\caption{\it 
Expected exclusion limits at 95\% C.L. by using the associated production 
of an electron with transverse missing energy, $pp\to e^\pm \slashed{E}_T$, 
at the LHC with center of mass energy
$\sqrt{s}=13\tev$ and a total luminosity $\call=139\fb^{-1}$.
}
\label{fig:95CL:mj:LHC13}
\end{figure}
Fig.~\ref{fig:95CL:mj:LHC13} shows the expected exclusion limits at 95\% C.L. in the (\(m_\chi\), \(\Lambda\)) plane.
The mono-electron process is most sensitive to (axial)-vector operators: 
for a massless dark fermion, 
the strongest lower limit reaches about \(12\ \text{TeV}\). 
Constraints on (pseudo)-scalar operators are slightly weaker, 
with the lower limit at approximately \(11.6\ \text{TeV}\) when \(m_\chi \approx 0\). 
The weakest bounds apply to tensor operators, 
yielding a lower limit of about \(8.4\ \text{TeV}\) for an ultra-light dark fermion 
(\(m_\chi \ll 1\ \text{TeV}\)).
Notably, these differences gradually diminish as the dark fermion mass increases. 
For a heavy dark fermion with \(m_\chi \sim 5\ \text{TeV}\), 
the lower bound on the energy scale is roughly \(2\ \text{TeV}\) 
for all operator types.

\subsection{Associated Production of Lepton with Photon}
\label{sec:LepPho}
\begin{figure}[th]
\centering
\includegraphics[height=0.12\textheight]{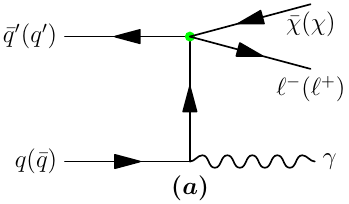}
\quad
\includegraphics[height=0.12\textheight]{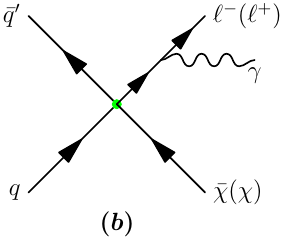}
\\[5mm]
\includegraphics[height=0.12\textheight]{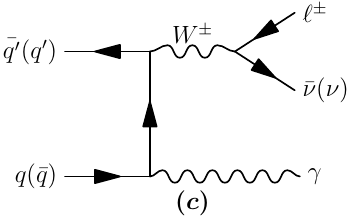}
\quad
\includegraphics[height=0.12\textheight]{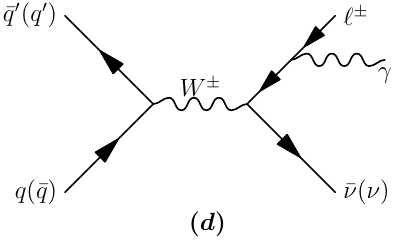}
\quad
\includegraphics[height=0.12\textheight]{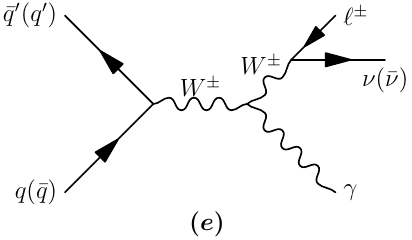}
\caption{\it 
Feynman diagrams of the associated production of a lepton and a photon
with missing energy: \textbf{(a)} and \textbf{(b)} are for the signal operators, 
\textbf{(c)}, \textbf{(d)} and \textbf{(e)} are for the irreducible background.
}
\label{fig:Feyn:LepPho}
\end{figure}

In addition to the mono-lepton signal, 
the effective operators defined in \eqref{eq:effo} can also induce 
the following associated production of transverse missing energy 
with a lepton and a photon,
\bee
p + p \;\to\; e^\pm + \gamma + \slashed{E}_T + X\,.
\ene
For signal processes, the photon can emerge from initial-state or final-state radiation, 
as shown in Fig.~\ref{fig:Feyn:LepPho} (a) and (b) for the corresponding 
parton-level Feynman diagrams, respectively. 
For backgrounds, besides photons emitted from incoming partons 
or outgoing leptons (as in Fig.~\ref{fig:Feyn:LepPho} (c) and (d)), 
an additional contribution arises from \(W^\pm\) radiation, 
as shown in Fig.~\ref{fig:Feyn:LepPho} (e).
Compared to mono-lepton production, 
both signal and background are suppressed by a factor of \(\alpha_{\text{EM}}\). 
Consequently, signal significance is suppressed by 
\(\sqrt{\alpha_{\text{EM}}}\) --- roughly an order of magnitude reduction. 
Although the cross section scales as \(\Lambda^{-4}\) 
(reducing the energy scale lower bound by a factor of $\sim 2$), 
the additional visible particles in the final state enable extraction of further information about the interaction vertex's Lorentz structure. 
Thus, detailed study of this channel is worthwhile.

\begin{figure}[th]
\centering
\includegraphics[width=0.44\textwidth]{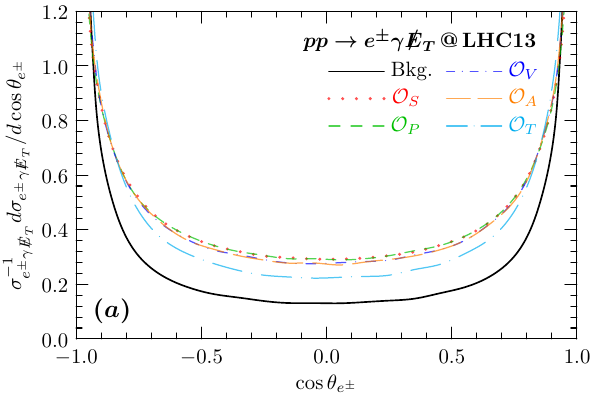}
\quad
\includegraphics[width=0.44\textwidth]{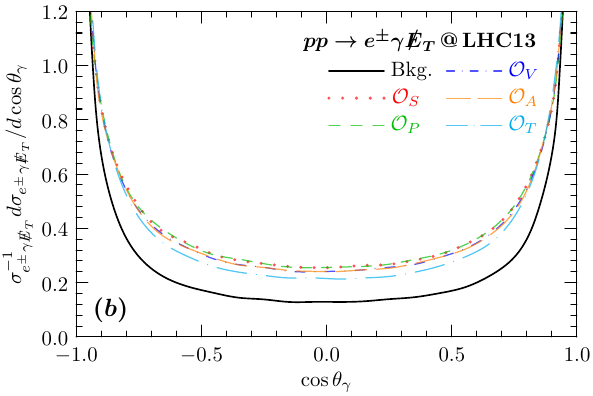}
\\
\includegraphics[width=0.44\textwidth]{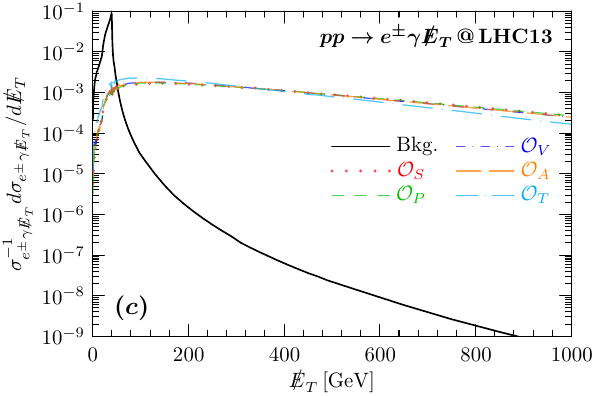}
\quad
\includegraphics[width=0.44\textwidth]{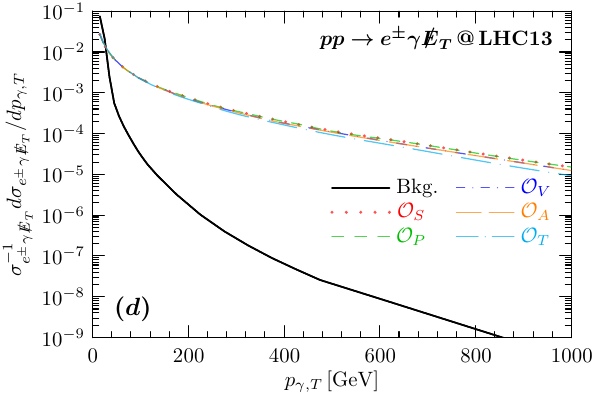}
\\
\includegraphics[width=0.44\textwidth]{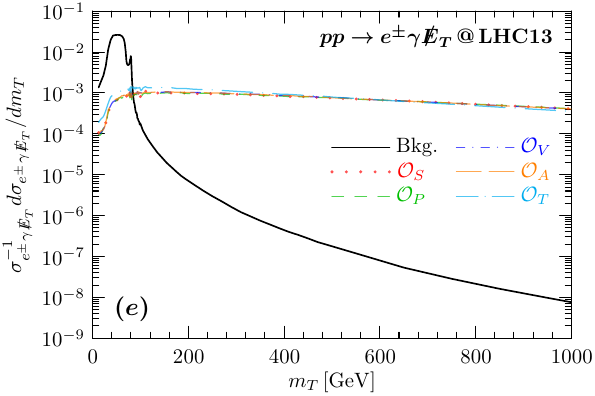}
\quad
\includegraphics[width=0.44\textwidth]{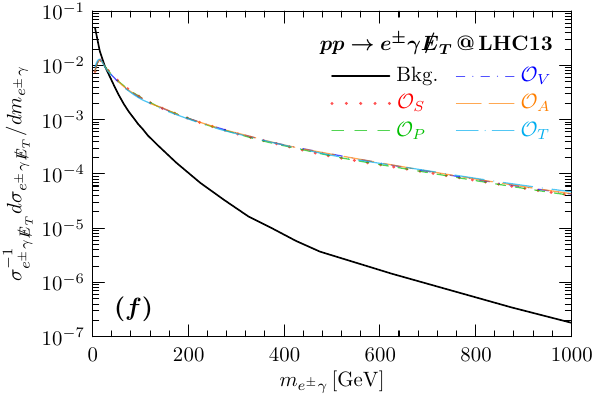}
\caption{\it 
Normalized distributions of the polar angles of the outgoing electron 
($\theta_{e^\pm}$) \textbf{(a)} and photon ($\theta_{\gamma}$) \textbf{(b)}, 
transverse momentum of the missing energy ($\slashed{E}_T$) \textbf{(c)}
and the photon ($p_{\gamma, T}$) \textbf{(d)},
transverse mass of the electron and the missing energy ($m_T$) \textbf{(e)} and 
invariant mass of the outgoing electron and photon ($m_{e^\pm\gamma}$) \textbf{(f)}
at the generator level in the laboratory frame with center of mass energy $\sqrt{s} = 13\tev$. 
In the above panels, the signals (colorful non-solid curves) are shown for parameters 
$m_\chi = 0\gev$ and $\varLambda_i = 1\tev$, 
and the background (black-solid curve) stands for the irreducible
contribution from the channel $pp \to \to e^\pm \gamma \nu$.
}
\label{fig:DXS:LAX:LHC13}
\end{figure}

Figs. \ref{fig:DXS:LAX:LHC13} (a) and (b) 
show normalized polar angle distributions of the electron (\(\theta_{e^\pm}\)) 
and photon (\(\theta_\gamma\)) 
in the laboratory frame at a center-of-mass energy of \(\sqrt{s} = 13\ \text{TeV}\), 
respectively. Signal event properties (colorful non-solid curves) are shown for 
parameters \(m_\chi = 0\ \text{GeV}\) and \(\Lambda_i = 1\ \text{TeV}\), 
with the irreducible background depicted by the black solid curve.
From Figs. \ref{fig:DXS:LAX:LHC13} (a), 
the electron polar angle distributions for both signal and irreducible background 
are highly similar to the mono-electron production case: 
most electrons are distributed in the forward and backward regions. 
Since photons originate from initial-state and final-state radiation, 
most photons also populate the forward and backward regions, 
as seen in Figs. \ref{fig:DXS:LAX:LHC13} (b).

The normalized distributions of missing transverse energy \(\slashed{E}_T\) 
and photon transverse momentum \(p_{T,\gamma}\) are shown in Fig. \ref{fig:DXS:LAX:LHC13} (c) and (d), respectively. 
For the irreducible background, 
since missing energy arises from \(W^{\pm}\) boson decay, 
\(\slashed{E}_T\) exhibits a sharp resonance peak 
at \(\slashed{E}_T \approx m_W/2\). In contrast, 
the photon transverse momentum distribution 
(Fig. \ref{fig:DXS:LAX:LHC13} (d)) lacks such a peak.
Fig. \ref{fig:DXS:LAX:LHC13} (e) shows the normalized distribution 
of the transverse mass \(m_T\) (defined in \eqref{eq:MT:LM}). 
Due to the presence of an additional photon, 
most events populate the low-\(m_T\) region, 
though a peak at \(m_T \approx m_W\) persists. 
Fig. \ref{fig:DXS:LAX:LHC13} (f) shows the normalized distribution 
of the electron-photon invariant mass \(m_{e^\pm\gamma}\). 
Both signal and background decrease as \(m_{e^\pm\gamma}\) increases, 
but the background drops more rapidly.
This feature --- stronger background suppression at high values --- 
is common to all four observables: \(\slashed{E}_T\), \(p_{T,\gamma}\), \(m_T\), and \(m_{e^\pm\gamma}\). 
Consequently, all can enhance signal significance.

\begin{figure}[th]
\centering
\includegraphics[width=0.44\textwidth]{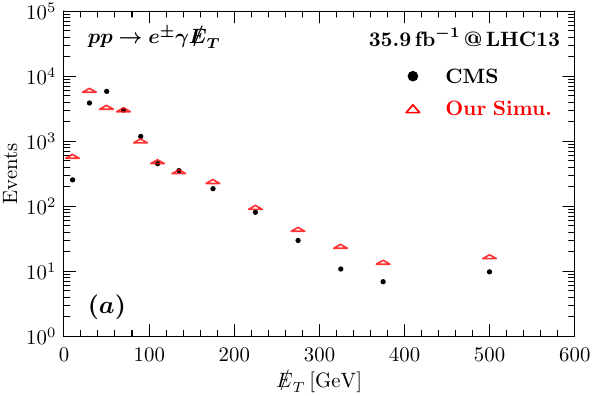}
\quad
\includegraphics[width=0.44\textwidth]{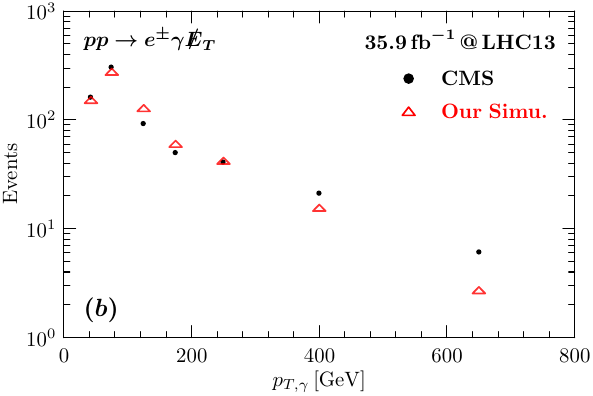}
\caption{\it 
Validation of our simulations for the transverse missing energy ($\slashed{E}_T$) 
and transverse momentum of the photon $p_{\gamma, T}$ distributions 
of the irreducible background process $pp \to e^\pm\gamma \nu$
at the LHC with center of mass energy $\sqrt{s} = 13\tev$ 
and a total luminosity $\call = 35.9\fb^{-1}$.
The experimental results (black dots) are taken from the Ref. \cite{CMS:2018fon},
and our results (red triangles) have been renormalized by multiplying 
an overall constant such that the total number of events is matched.
}
\label{fig:V:LAX:LHC13}
\end{figure}

The CMS collaboration searched for supersymmetry in events with a photon, 
a charged lepton, and large missing transverse momentum \cite{CMS:2018fon} 
at the LHC with a center-of-mass energy of \(\sqrt{s} = 13\ \text{TeV}\) 
and total luminosity of \(35.9\ \text{fb}^{-1}\). 
Here, we use the CMS's result as a reference dataset to validate our simulation.
Events are selected with the following requirements: 
1) Photon transverse momentum \(p_{T,\gamma} > 35\ \text{GeV}\) 
and rapidity \(|\eta_\gamma| < 1.44\); 
2) Electron transverse momentum \(p_{T,e} > 25\ \text{GeV}\), 
rapidity \(|\eta_e| < 2.5\) and \(|\eta_e| \notin [1.44, 1.56]\) 
(to exclude the barrel-endcap transition region); 
3) Large angular separation between the photon and electron, 
\(\Delta R_{e^\pm\gamma} > 0.5\), to suppress final-state radiation contributions;
4) Invariant mass \(m_{e^\pm\gamma}\) at least \(10\ \text{GeV}\) above the 
\(Z\) boson mass, reducing contamination from \(Z\to e^+e^-\) where one electron is misidentified as a photon.
The signal region is defined by \(\slashed{E}_T > 120\ \text{GeV}\) 
and \(m_T > 100\ \text{GeV}\). 
Due to the absence of parton shower effects in our simulation, 
we applied a slightly smaller but reasonable cut on transverse mass, 
\(m_T > 80\ \text{GeV}\), to avoid excessive background reduction. 
Figs. \ref{fig:V:LAX:LHC13} (a) and (b) show validations of our simulations 
for the irreducible background process \(pp \to e^\pm \gamma \nu\), 
for the observables \(\slashed{E}_T\) and \(p_{T,\gamma}\).
Our result is renormalized by a universal scale factor 
to account for the strong efficiency reduction caused 
by the cut \(m_T > 80\ \text{GeV}\) and potential detector effects. 
One can see there are some difference between 
our results and the experimental results.
However, since the cross section scales as \(\varLambda^{-4}\), 
the overall normalization factor (\(\alpha\)) affects the exclusion limit 
by a factor of $(\sqrt{\alpha})^{1/4}$. This is a very mild effect: 
for \(\alpha=2\), the exclusion limit scales by approximately 1.1, 
and for \(\alpha=10\), it scales by $\sim 1.3$. 
Therefore, for both observables, CMS's results are reasonably reproduced. 
A moderate discrepancy exists in the soft region of missing transverse energy, 
but this does not affect our exclusion limit estimation due to the strict 
\(\slashed{E}_T > 120\ \text{GeV}\) cut. 
Thus, the approximation of an overall normalization factor effectively 
captures both total event counts and differential distributions.

The CMS collaboration used a sophisticated procedure to estimate the exclusion limit.
By contrast, given our focus on a different parameter space of interest, the expected exclusion limits are estimated by calculating the following $\chi^2$,
\bee
\label{eq:chi2:ndata}
\chi^2  =  \sum_{\calo} \sum_{ \calo_i}
 \frac{\left[  \epsilon_D \cdot N^{\rm S}_{ \calo_i }  \right]^2}
 { \sqrt{ \epsilon_D \cdot \left( \epsilon_{R} \cdot N^{B}_{ \calo_i } + N^{\rm S}_{ \calo_i } \right) } }\,,
\ene
where \(\calo \in \{\slashed{E}_T, p_{T,\gamma}, m_T, m_{e^\pm\gamma}\}\) 
denote the four observables used to enhance signal significance. 
For the $i$-th bin of observable \(\calo\), \(N^{\rm S}_{\calo,i}\) 
and \(N^{\rm B}_{\calo,i}\) are the numbers of signal and irreducible background events, respectively. 
\(\epsilon_D\) is a universal normalization factor accounting for detector efficiency, 
as discussed earlier, and the scale factor \(\epsilon_R\) is used to account for reducible background contributions.

\begin{figure}[h]
\centering
\includegraphics[width=0.48\textwidth]{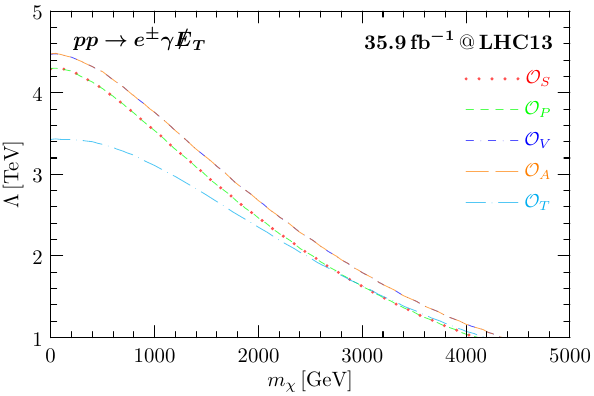}
\caption{\it 
Expected exclusion limits at 95\% C.L. by using the associated production 
of an electron and a photon with transverse missing energy,
$pp\to e^\pm \gamma \slashed{E}_T$, 
at the LHC with center of mass energy $\sqrt{s}=13\tev$ 
and a total luminosity $\call=35.9\fb^{-1}$.
}
\label{fig:EL:LAX:LHC13}
\end{figure}
Fig.~\ref{fig:EL:LAX:LHC13} shows the expected exclusion limits at 95\% C.L. 
in the (\(m_\chi\), \(\Lambda\)) plane. 
As expected (similar to mono-electron production), 
constraints on (axial)-vector operators are strongest, 
followed by slightly weaker constraints on (pseudo)-scalar operators, 
with tensor operators yielding the weakest bounds. 
For a massless dark fermion, the lower bounds on the energy scale for 
(axial)-vector, (pseudo)-scalar, and tensor operators reach about 
\(4.43\ \text{TeV}\), \(4.25\ \text{TeV}\), and \(3.41\ \text{TeV}\), respectively.
Notably, these differences gradually diminish for heavy dark fermions. 
For \(\Lambda_i = 1\ \text{TeV}\), 
dark fermions with masses up to approximately \(4\ \text{TeV}\) 
can be excluded for all operator types.

\subsection{Associated Production of Lepton with Jet}
\label{sec:LepJet}
\begin{figure}[h]
\centering
\includegraphics[height=0.11\textheight]{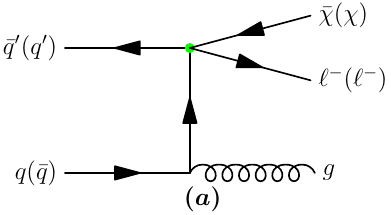}
\;
\includegraphics[height=0.11\textheight]{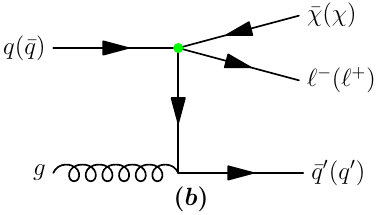}
\;
\includegraphics[height=0.11\textheight]{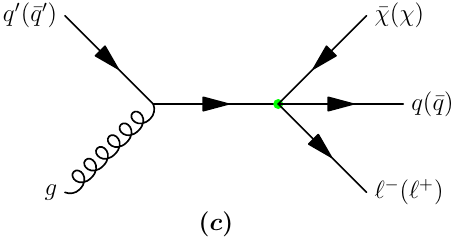}
\\[5mm]
\includegraphics[height=0.12\textheight]{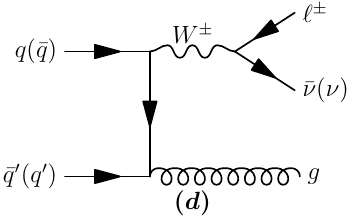}
\;
\includegraphics[height=0.12\textheight]{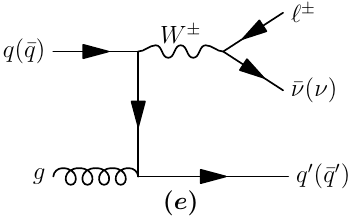}
\;
\includegraphics[height=0.12\textheight]{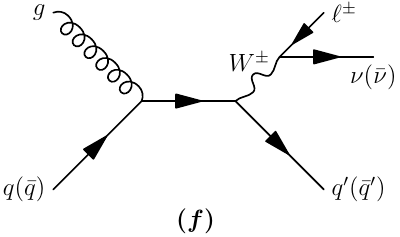}
\caption{\it 
Feynman diagrams of the associated production of a lepton and one jet
with missing energy: 
\textbf{(a)}, \textbf{(b)} and \textbf{(c)} are for the signal operators, 
\textbf{(d)}, \textbf{(e)} and \textbf{(f)} are for the irreducible background.
}
\label{fig:Feyn:LepJet}
\end{figure}

Similar to the associated production of missing transverse energy 
with a lepton and a photon through initial and final state radiations, 
the effective operators defined in \eqref{eq:effo} can also induce 
the following associated production of missing transverse energy 
with a lepton and one jet,
\begin{equation}
p + p \;\to\; e^\pm + j + \slashed{E}_T + X\,.
\end{equation}
However, in this case the additional jet in the signal process 
can not only be generated from emission of incoming partons 
(as shown in Fig.~\ref{fig:Feyn:LepJet} (a) and (b) 
for the parton-level Feynman diagrams), 
but also can be generated via an $s$-channel exchange of an off-shell quark, 
as shown in Fig.~\ref{fig:Feyn:LepJet} (c). Furthermore, 
the gluons of the incoming hadrons can give nontrivial contributions. 
This is also true for the irreducible background, 
as shown in Fig.~\ref{fig:Feyn:LepJet} (d), (e) and (f).
Compared to the mono-electron production channel, 
the total cross sections of both the signal and background processes 
of this associated production are suppressed by a factor of $\alpha_{\rm s}$. 
As a result, the signal significance is reduced by a factor of about 
$\sqrt{\alpha_{\rm s}}$, roughly a third reduction. 
For the energy scale $\varLambda_i$, the suppression factor is about $3/4$. 
Hence, it is expected to have a comparable exclusion limit 
to the mono-electron production channel. Again, 
because there is one more visible particle in the final state, 
more information about the Lorentz structures of the interaction vertex 
can be obtained by further analysis of the angular distributions and 
correlations of the final state particles. Furthermore, 
since the initial state configuration is completely different from 
that of the associated production of missing energy with a lepton and a photon, 
and the generation channels are also different, 
it is expected that sensitivities to different operators are also different. 
Hence, it is worthy to study this production process in detail.

\begin{figure}[th]
\centering
\includegraphics[width=0.44\textwidth]{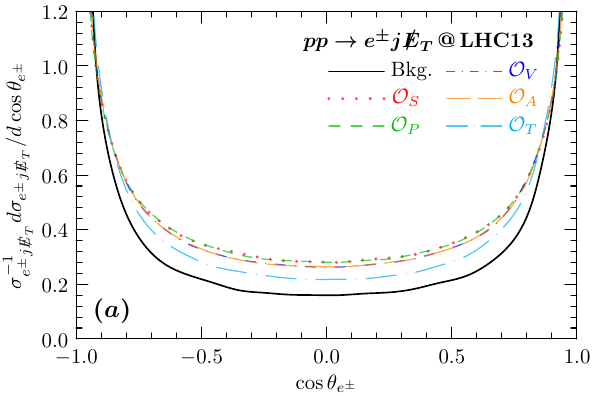}
\quad
\includegraphics[width=0.44\textwidth]{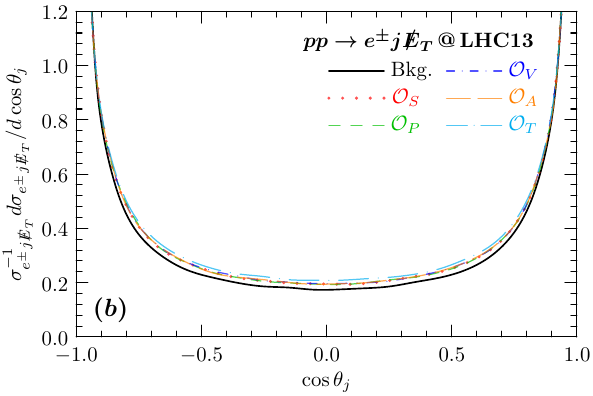}
\\
\includegraphics[width=0.44\textwidth]{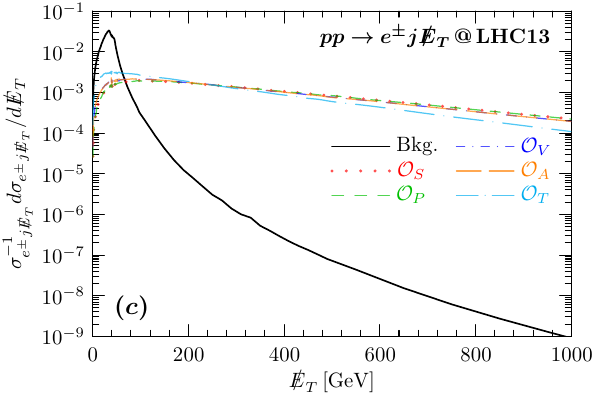}
\quad
\includegraphics[width=0.44\textwidth]{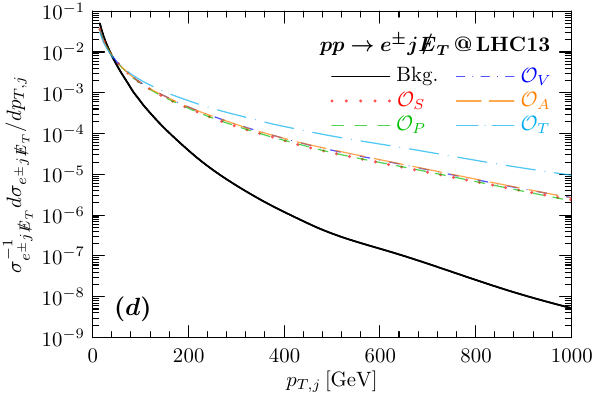}
\\
\includegraphics[width=0.44\textwidth]{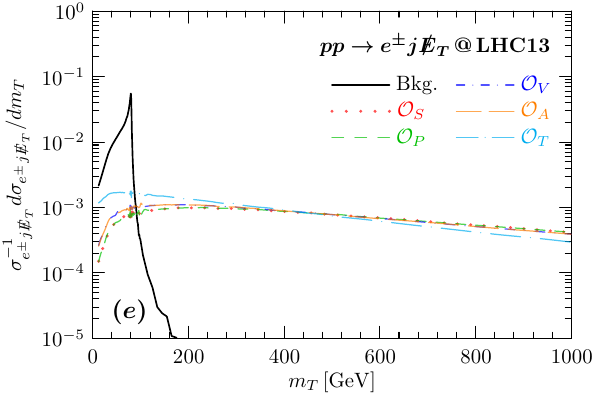}
\quad
\includegraphics[width=0.44\textwidth]{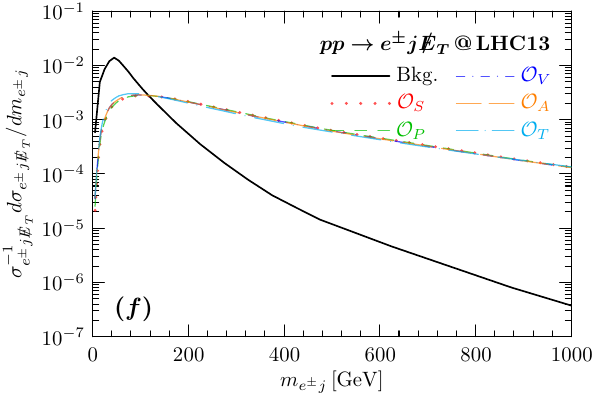}
\caption{\it 
Normalized distributions of the polar angles of the outgoing electron 
($\theta_{e^\pm}$) \textbf{(a)} and the jet ($\theta_{j}$) \textbf{(b)}, 
transverse momentum of the missing energy ($\slashed{E}_T$) \textbf{(c)}
and the jet ($p_{j, T}$) \textbf{(d)},
transverse mass of the electron and the missing energy ($m_T$) \textbf{(e)} and 
invariant mass of the outgoing electron and jet ($m_{e^\pm j}$) \textbf{(f)}
at the generator level in the laboratory frame with center of mass energy $\sqrt{s} = 13\tev$. 
In the above panels, the signals (colorful non-solid curves) are shown for parameters 
$m_\chi = 0\gev$ and $\varLambda_i = 1\tev$, 
and the background (black-solid curve) stands for the irreducible
contribution from the channel $pp \to \to e^\pm j \nu$.
}
\label{fig:DXS:LJX:LHC13}
\end{figure}

Fig.~\ref{fig:DXS:LJX:LHC13} (a) and (b) show the normalized 
polar angle distributions of the electron ($\theta_{e^\pm}$) and 
of the jet ($\theta_{\mathrm{j}}$) in the laboratory frame 
at a center-of-mass energy of $\sqrt{s}=13\,\text{TeV}$, respectively. 
The distributions of the signal events (colorful non-solid curves) are shown 
for parameters $m_\chi = 0\,\text{GeV}$ (dark matter mass) and 
$\varLambda_i = 1\,\text{TeV}$, while the irreducible background is shown 
by the black solid curve. One can clearly see that 
the polar angle distributions of the electron and jet are dominant 
in the forward and backward regions. 
This is simply because the major production mechanism 
involves initial state radiation  (of $W^\pm$ boson and jet).

The normalized distributions of the missing transverse energy 
($\slashed{E}_T$) and transverse momentum of the jet ($p_{j, T}$) 
are shown in Fig.~\ref{fig:DXS:LJX:LHC13} (c) and (d), respectively. 
Again, since the transverse missing energy of the irreducible background 
comes from the neutrino of the $W^{\pm}$ boson decay, 
its distribution has a peak at $\slashed{E}_T \sim m_W/2 \sim 40\,\text{GeV}$. 
However, the peak is not as sharp as the peaks in the mono-electron production 
and the associated production of missing energy with an electron and a photon. 
This is because when the incoming parton is a gluon, 
the radiated parton can carry more energy than the photon 
in similar initial-state radiation processes. 
This can be seen by comparing Fig.~\ref{fig:DXS:LJX:LHC13} (d) 
and Fig.~\ref{fig:DXS:LAX:LHC13} (d): 
the transverse momentum of the jet drops more slowly than that of the photon.  

Fig.~\ref{fig:DXS:LJX:LHC13} (e) shows the normalized distribution 
of the transverse mass of the electron and missing momentum 
(as defined in \eqref{eq:MT:LM}). Since the emitted jet is completely 
isolated from the $W^\pm$ boson decay, 
the constraint on the transverse mass $m_T$ is comparable to that 
in the mono-electron production channel 
(stronger than the associated production of missing energy with $e^\pm$ and 
$\gamma$, as shown in Fig.~\ref{fig:DXS:LAX:LHC13} (e)). 
One can see a very sharp peak at $m_T \sim m_W \sim 80\,\text{GeV}$.  

Fig.~\ref{fig:DXS:LJX:LHC13} (f) shows the normalized distribution 
of the invariant mass ($m_{e^\pm j}$) of the outgoing electron and jet. 
The background events peak at low values (soft region), 
while the signal contributions remain sizable at high invariant masses. 
This feature --- background events populating low values and 
signal events dominating at high values --- 
holds for all four observables: transverse missing energy $\slashed{E}_T$, 
jet transverse momentum $p_{j, T}$, transverse mass $m_T$, 
and invariant mass $m_{e^\pm j}$. In our estimation of the exclusion limits, 
all of them will be used to enhance the signal significance.

The ATLAS collaboration searched for quantum black hole production 
in the single-lepton plus jet final state at the LHC with 
a centre-of-mass energy $\sqrt{s}=13\tev$ and 
an integrated luminosity of $140\fb^{-1}$ \cite{ATLAS:2023vat}. 
Here we take ATLAS’s result as the reference data set to validate our simulation.  
The events are selected by requiring that:  
1) the jet transverse momentum $p_{j,T} > 130\gev$, with rapidity $|\eta_j| < 2.8$;  
2) the electron transverse momentum $p_{e,T} > 130\gev$, with rapidity restricted to $|\eta_e| < 2.5$ and $|\eta_e| \notin [1.37, 1.52]$ to exclude the barrel-endcap transition region;  
3) the jet is separated from the electron by a distance $\Delta R_{e^\pm j} > 0.4$.  
Furthermore, the invariant mass $m_{e^\pm j}$ is required to be 
at least $10\gev$ above the $Z$ boson mass to reduce contamination 
from $Z\to e^+e^-$ events where one electron is misidentified as a jet. 
The signal region is defined by $m_{e^\pm j} > 1\tev$.
\begin{figure}[th]
\centering
\includegraphics[width=0.48\textwidth]{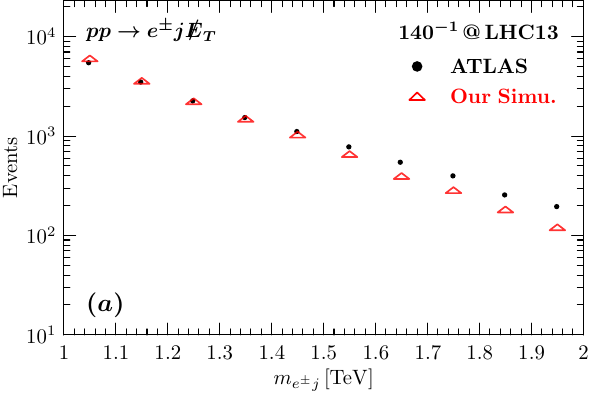}
\includegraphics[width=0.48\textwidth]{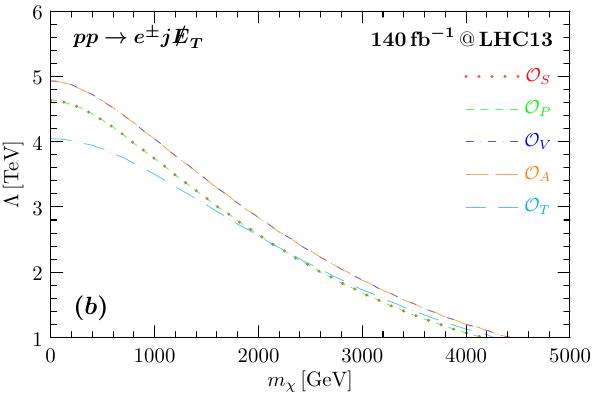}
\caption{\it 
\textbf{(a)}: 
Validation of our simulation for the invariant mass ($m_{e^\pm j}$) distribution of 
the outgoing electron and jet of the irreducible background process 
$pp \to e^\pm j \nu$ at the LHC with center of mass energy $\sqrt{s} = 13\tev$ 
and a total luminosity $\call = 140\fb^{-1}$.
The experimental results (black dots) are taken from the Ref. \cite{ATLAS:2023vat},
and our results (red triangles) have been renormalized by multiplying 
an overall constant such that the total number of events is matched.
\textbf{(b)}: 
Expected exclusion limits at 95\% C.L. by using the validated simulation 
of the process $pp\to e^\pm j \slashed{E}_T$
at the LHC with center of mass energy $\sqrt{s}=13\tev$ 
and a total luminosity $\call=35.9\fb^{-1}$.
}
\label{fig:V:LJX:LHC13}
\end{figure}

Fig.~\ref{fig:V:LJX:LHC13} (a) shows the validation of our simulation 
for the irreducible background process $pp \to e^\pm j \nu$ 
(where $\nu$ is a neutrino), presenting the invariant mass distribution of 
the electron and jet, $m_{e^\pm j}$. Our result is renormalized 
by a global normalization factor $\epsilon_D = 0.058$ 
(extracted by matching the total event number of our simulation to that 
reported in the ATLAS paper \cite{ATLAS:2023vat}). 
This factor accounts for detector effects such as efficiency and resolution. 
One can see that our simulation result shows good agreement 
with the ATLAS data. Hence, the global normalization approximation 
works well for both the total number of events and the differential distributions.

For estimating the exclusion limit, 
since we are interested in a different parameter space, 
we define the signal region by the condition $m_T > 100\gev$ 
(instead of requiring $m_{e^\pm j} > 1\tev$). 
As seen in Fig.~\ref{fig:DXS:LJX:LHC13} (e), 
the transverse mass is a powerful observable for background suppression. 
Within this signal region, the expected exclusion limits are estimated 
by calculating the $\chi^2$ function defined in \eqref{eq:chi2:ndata}.  

Fig.~\ref{fig:V:LJX:LHC13} (b) shows the 95\% expected exclusion limits 
in the $m_\chi$–$\Lambda$ plane for our signal operators. 
Similar to the mono-electron and $e^\pm\gamma$ associated production, 
the strongest bounds arise for (axial-)vector operators. 
The key difference is that the constraint on the tensor operator is comparable to, 
or even stronger than, those on (pseudo-)scalar operators. 
This is mainly due to the different mass dependence of the cross sections for tensor and (pseudo-)scalar operators. 
Such a difference can only be manifested when \(m_\chi \sim \sqrt{\hat{s}}\).
For a massless dark fermion, the lower bounds on the energy scales 
of (axial-)vector, (pseudo-)scalar, and tensor operators reach about 
$4.4\tev$, $4.0\tev$, and $4.0\tev$, respectively. 
Notably, these differences gradually diminish for a heavy dark fermion. 
For $\Lambda_i = 1\tev$, a dark fermion with mass up to approximately 
$4\tev$ can be excluded for all operator types.

\subsection{Constraints at Future Hadron Colliders}
\label{sec:Future}
As a general property, 
the signal cross sections of the four-fermion contact operators 
grow quickly with respect to the centre-of-mass energy, 
while that of the corresponding background decreases. 
Thus, it is worthwhile to study future sensitivities 
for the upgraded LHC \cite{CidVidal:2018eel}. Furthermore, 
the total luminosity accumulated at the upgrades will be much larger 
than that of the current LHC, 
giving the upgraded LHC great advantages for probing such signals.  
Tab. \ref{tab:hptc} lists the high-luminosity LHC configurations studied in this paper:  
1) LHC14: centre-of-mass energy $\sqrt{s}=14\,\text{TeV}$ and integrated luminosity $\mathcal{L}=3\,\text{ab}^{-1}$;  
2) LHC25: centre-of-mass energy $\sqrt{s}=25\,\text{TeV}$ and integrated luminosity $\mathcal{L}=20\,\text{ab}^{-1}$.  
The corresponding kinematic cuts for the three production processes are also listed in Tab. \ref{tab:hptc}.
\begin{table}[th]
\renewcommand\arraystretch{1.44}
\begin{center}
\begin{tabular}{c c c c}
{\rm Process}   & 14~TeV, 3\iab & 25~TeV, 20\iab
%
%
\\\hline\hline
$pp\to e^\pm \met$ 
& 
\begin{tabular}{c} 
$p_{T,e} \geqslant 65\gev$\,,
$\big|\eta_{e}\big| \leqslant 2.5$
\\ 
$ \met \geqslant 65\gev$ 
\end{tabular}
& 
\begin{tabular}{c} 
$p_{T,e} \geqslant 150\gev$\,,
$\big|\eta_{e}\big| \leqslant 2.5$
\\ 
$ \met \geqslant 150\,\gev$ 
\end{tabular}
\\[5mm]\hline
$pp\to e^\pm\gamma\met$ 
& 
\begin{tabular}{c} 
$p_{T,e} \geqslant 25\gev$, $\big|\eta_{e}\big| \leqslant 2.5$ 
\\
$p_{T,\gamma} \geqslant 35\gev$, $\big|\eta_{\gamma}\big| \leqslant 1.44$  
\\
$\Delta R_{e, \gamma} \geqslant 0.8$
\\
$|m(e, \gamma) - m_Z | > 10\gev$
\\
$\met \geqslant 120\gev$, $m_T \geqslant 80\gev$
\end{tabular}
& 
\begin{tabular}{c} 
$p_{T,e} \geqslant 50\gev$, $\big|\eta_{e}\big| \leqslant 2.5$ 
\\
$p_{T,\gamma} \geqslant 70\gev$, $\big|\eta_{\gamma}\big| \leqslant 1.44$  
\\
$\Delta R_{e, \gamma} \geqslant 1$
\\
$|m(e, \gamma) - m_Z | > 10\gev$
\\
$\met \geqslant 200\gev$, $m_T \geqslant 80\gev$
\end{tabular}
\\[5mm]\hline
$pp\to e^\pm j\met$ 
& 
\begin{tabular}{c} 
$p_{T,e} \geqslant 130\gev$, $\big|\eta_{e}\big| \leqslant 2.5$ 
\\
$p_{T, j} \geqslant 130\gev$, $\big|\eta_{j}\big| \leqslant 2.8$  
\\
$\Delta R_{e, j} \geqslant 0.4$
\\
$|m(e, j) - m_Z | > 10\gev$
\\
$m_T \geqslant 100\gev$
\end{tabular}
& 
\begin{tabular}{c} 
$p_{T,e} \geqslant 250\gev$, $\big|\eta_{e}\big| \leqslant 2.5$ 
\\
$p_{T, j} \geqslant 250\gev$, $\big|\eta_{j}\big| \leqslant 2.8$  
\\
$\Delta R_{e, j} \geqslant 0.4$
\\
$|m(e, j) - m_Z | > 10\gev$
\\
$m_T \geqslant 200\gev$
\end{tabular}
\\[2mm]\hline\hline
\end{tabular}
\caption{\it Configurations of the upgrades of the LHC and the corresponding kinematical cuts at parton level.}
\label{tab:hptc}
\end{center}
\end{table}

\begin{figure}[h]
\centering
\includegraphics[width=0.32\textwidth]{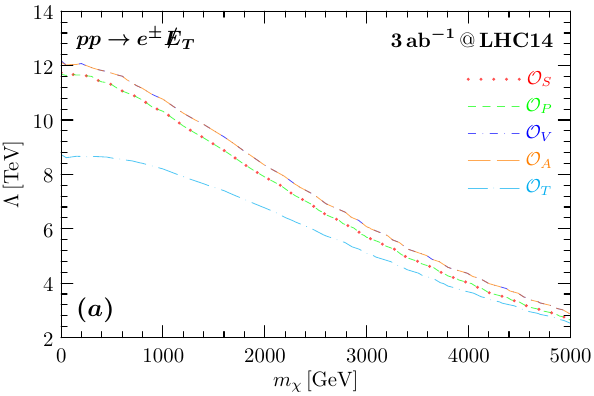}
\includegraphics[width=0.32\textwidth]{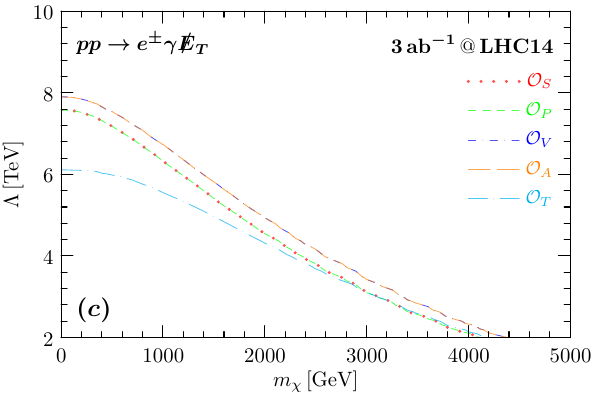}
\includegraphics[width=0.32\textwidth]{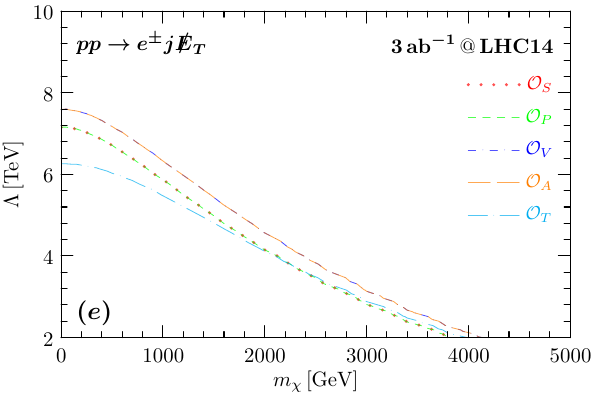}
\\
\includegraphics[width=0.32\textwidth]{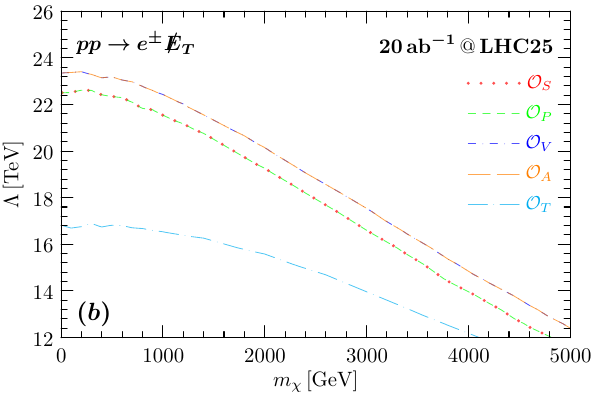}
\includegraphics[width=0.32\textwidth]{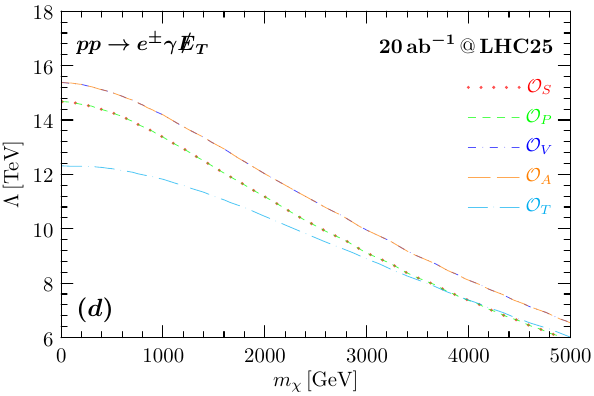}
\includegraphics[width=0.32\textwidth]{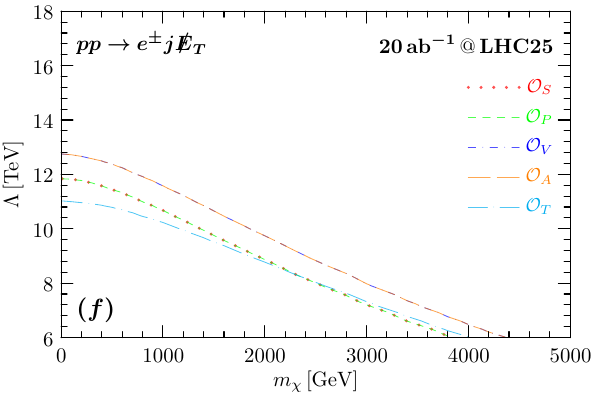}
\caption{\it 
Expected exclusion limits at 95\% C.L.  at the LHC 
with center of mass energy $\sqrt{s}=14\tev$ and a total luminosity 
$\call=3\ab^{-1}$ (\textbf{top panels})
, and $\sqrt{s}=25\tev$ and a total luminosity 
$\call=20\ab^{-1}$ (\textbf{bottom pannels}):
\textbf{(a)} and \textbf{(b)} stands for the mono-electron production channel,
\textbf{(c)} and \textbf{(d)} stands for the associated production of an electron
and a photon with missing energy,
\textbf{(e)} and \textbf{(f)} stands for the associated production of an electron
and one jet with missing energy.
}
\label{fig:C:Fut:LHC}
\end{figure}
Fig.~\ref{fig:C:Fut:LHC} shows the expected exclusion limits at 95\% C.L. 
for LHC14 (top panels) and $\sqrt{s}=25\tev$ (bottom panels). 
From left to right, Fig.~\ref{fig:C:Fut:LHC} (a) and (b) 
stand for the mono-electron production channel, 
Fig.~\ref{fig:C:Fut:LHC} (c) and (d) show results 
for the associated production of an electron and a photon with missing energy, 
and Fig.~\ref{fig:C:Fut:LHC} (e) and (f) present results 
for the associated production of an electron and a jet with missing energy.  
One can see that mono-electron production remains the most sensitive 
process to four-fermion contact operators. 
The $e^\pm\gamma$ associated production gives much weaker constraints, 
but slightly stronger than those from the $e^\pm j$ associated production. 
Furthermore, the strongest constraints arise for (axial-)vector operators, 
followed by slightly weaker bounds on (pseudo-)scalar operators, 
with the weakest constraints on tensor operators --- 
a trend common to all three channels.  

For the two associated production channels, 
LHC14 enhances constraints by a factor of about 1.5 $\sim$ 1.7, 
whereas the enhancement in the mono-electron channel is negligible. 
This may be because we used identical bins (extracted from the ATLAS paper) 
for both LHC13 and LHC14, excluding some phase space regions with higher 
signal significance. In contrast, LHC25 achieves a factor of 2 enhancement. 
In the next section, we discuss how these improvements 
interplay with direction detection experiments.

\section{Induced $\beta$ Decay}
\label{sec:InvBetaDecay}
The effective operators defined in \eqref{eq:effo} can also lead to 
nontrivial signals in direct detection experiments \cite{Dror:2019onn,Dror:2019dib}. 
While the signal of the four-fermion operator involving a dark fermion 
and a neutrino appears as nuclear recoil, 
the operator involving a dark fermion and a charged lepton 
can induce $\beta$-decay. 
If the dark fermion mass $m_\chi$ exceeds the threshold, 
\(
m_{\mathrm{th}}^{\beta^{\mp}} \equiv M_{A, Z \pm 1}^{*} + m_{e} - M_{A, Z}
\)
(where $M_{A, Z}$ is the mass of the nucleus ${}_{Z}^{A}\mathrm{X}$), for the following two nuclear transitions,
\begin{equation}
\chi\left(\vec{p}_{\chi} =m_{\chi} \vec{v}_\chi\right) + { }_{Z}^{A} \mathrm{X}(\vec{p}_{i}=\vec{0}) 
\;\rightarrow\; 
{ }_{Z\pm1}^{A} \mathrm{X}(\vec{p}_{f}) + e^{\mp}\left(\vec{p}_{e}\right) \,,
\end{equation}
then, an induced $\beta$-decay transition can be triggered. 
The corresponding two absorption processes at the nucleon level are given as follows,
\begin{equation}
\begin{array}{l}
\beta^{-}: \quad \chi+n \;\rightarrow\; p+e^{-}  \,,
\\[3mm] 
\beta^{+}: \quad \overline{\chi}+p \;\rightarrow\;  n+e^{+}  \,.
\end{array}
\end{equation}
We will follow the approaches given in the Ref. \cite{Dror:2019dib} to study 
the above processes for constraints on the four fermion contact operators.

If the mass of the dark fermion is larger than the typical binding energy 
of the nucleus ($10\,\text{MeV}$), the outgoing proton/neutron after absorption 
can gain enough energy to escape the nucleus, 
and hence new signals can appear \cite{Formaggio:2012cpf}. 
Here we restrict ourselves to the case where $m_\chi < 10\,\text{MeV}$. 
In this case, the nuclei are seen by the dark fermion as a whole target, 
and the scattering rate is given as \cite{Dror:2019dib},
\begin{equation}
R = \frac{\rho_{\chi}}{2 m_{\chi}} \sum_{j} N_{T,j} n_j \langle\sigma v\rangle_j \,,
\end{equation}
where $\rho_\chi \approx 0.3\,\text{GeV}\cdot\text{cm}^{-3}$ 
is the local dark matter density; $j$ denotes the set of target isotopes, 
with $N_{T,j}$ as the number of targets for isotope $j$; 
$n_j$ is the total number of nucleons involved in the transition 
($n_j = A_j-Z_j$ for $\beta^-$ and $Z_j$ for $\beta^+$); 
and $\langle\sigma v\rangle_j$ is the thermally averaged scattering 
cross section for isotope $j$.

In the center-of-mass frame, the differential cross section is given as,
\begin{equation}
\frac{d \sigma}{d \Omega} = \frac{1}{64 \pi^{2} E_{\mathrm{cm}}^{2}} \frac{|\vec{p}_{e}|}{|\vec{p}_{\chi}|} \sum_{\text{transitions}} \overline{|\mathcal{M}_{N}|^{2}} \,,
\end{equation}
where $\mathcal{M}_N$ is the scattering amplitude, 
$d\Omega = d\cos\theta\,d\phi$ is the solid angle element of the emitted lepton, 
$E_{\mathrm{cm}}$ is the center-of-mass energy, 
and $\vec{p}_e$ and $\vec{p}_\chi = m_\chi \vec{v}_\chi$ are 
the momenta of the lepton and dark fermion, respectively. 
All nuclear spin states are summed over.  

For $m_e, m_\chi, m_{\mathrm{th},j}^\beta \ll M_{A_j,Z_j}$, 
the center-of-mass frame approximates the lab frame, 
with $E_{\mathrm{cm}} \approx M_{A_j,Z_j}$. 
The lepton momentum is approximately  
\(
|\vec{p}_e|_j^2 = (m_{\mathrm{th},j}^\beta - m_\chi)^2 - m_e^2 \,.
\)
In most cases, the scattering amplitude has trivial angular dependence 
(especially for vanishing momentum transfer), 
so the solid angle integrates to $4\pi$. 
The thermally averaged cross section then becomes  
\begin{equation}
\langle\sigma v\rangle_j = \frac{|\vec{p}_{e}|_j}{16 \pi m_{\chi} M_{A_j,Z_j}^{2}} \overline{|\mathcal{M}_{N_j}|^{2}} \,.
\end{equation}
Thus, the total transition rate can be written as,
\begin{equation}
R = \frac{\rho_{\chi}}{2 m_{\chi}} \sum_{j} \frac{N_{T,j} n_j |\vec{p}_{e}|_j}{16 \pi m_{\chi} M_{A_j,Z_j}^{2}} \overline{|\mathcal{M}_{N_j}|^{2}} \,.
\end{equation}

\subsection{Induced $\beta^-$ Decay}
\label{sec:IBD:IBMD}
The transition amplitude $\mathcal{M}_{N_j}$ can be simplified 
by employing the corresponding matrix element 
for the induced beta decay of the nucleus by a SM neutrino. 
In terms of the scattering amplitude $\mathcal{M}$ 
for the induced $\beta^-$ transition in a single nucleon 
(\textit{i.e.}, the process $\chi + n \to p + e^-$), 
the nuclear-level amplitude $\mathcal{M}_{N_j}$ can be factorized as \cite{Formaggio:2012cpf,Vogel:1999zy},
\begin{equation}
\mathcal{M}_{N} = \sqrt{\mathcal{F}(Z+1, E_e)} \, \mathcal{M} \,,
\end{equation}
where the factor $\mathcal{F}(Z+1, E_e)$ is the usual Fermi function, 
accounting for Coulomb interactions 
between the charged outgoing nucleon and electron. 
It is given by
\begin{equation}
\mathcal{F}(Z, E_e) = 2(1+S) \frac{|\Gamma(S+i \eta)|^2}{\Gamma(1+S)^2} \left(2 r_N |\vec{p}_e|\right)^{2S-2} e^{\pi \eta} \,,
\end{equation}
with $\eta = \alpha Z / \beta_e$ ($\beta_e = v_e/c$ being the electron velocity) 
and $S = \sqrt{1 - \alpha^2 Z^2}$. The nuclear radius is 
$r_N = 1.2\,\text{fm} \cdot A^{1/3}$. 
For energetic electrons ($E_e \gg m_e$), 
the Fermi function increases with proton number $Z$.  
Notably, the momentum of the nucleon in the process $\chi + n \to p + e^-$ should be normalized to the nuclear mass, 
\textit{i.e.}, $p^2_{\rm nucleon} = M_{A_j,Z_j}^2$.
The total event rate for the induced $\beta^-$ decay can be simply rewritten as,
\begin{equation}
R
=
\frac{\rho_{\chi}}{2 m_{\chi}} \sum_{j} 
\frac{ N_{T, j} n_{j}  \left|\vec{p}_{e}\right|_{j}}{16 \pi m_{\chi} M_{A_{j}, Z_{j}}^{2}} 
\mathcal{F}\left(Z+1, E_{e}\right) 
\overline{|\mathcal{M}|^{2}} \,.
\end{equation}

Calculation of the nucleon-level transition amplitude $\mathcal{M}$ 
is not straightforward starting from our definitions 
of the effective operators given in \eqref{eq:effo}. 
The matrix elements at the quark-level must be matched to 
those involving nucleons \cite{Bishara:2017pfq,DelNobile:2021wmp}, 
as the operators are defined at the quark level. 
The operators given in \eqref{eq:effo} are generally parameterized 
for all five possible Lorentz structures. For neutral currents, 
all four-fermion contact operators have non-vanishing low-energy counterparts  
\cite{Bishara:2017pfq,DelNobile:2021wmp}, 
whereas this is not true for charged currents.  

In standard beta decay, it is known that scalar and tensor couplings 
cannot be distinguished from vector and axial-vector couplings 
by the electron spectrum alone \cite{Lee:1956qn}. 
Additional contributions beyond the standard V-A electroweak coupling 
can only be probed via Fierz interference effects 
\cite{Jackson:1957zz,Avila-Aoki:1990fxk,Langacker:1991zr}. 
However, there is no evidence of non-vanishing contributions from scalar, 
pseudo-scalar, or tensor operators \cite{Anholm:2023xrb}, 
and Fierz interference is tightly constrained \cite{Beck:2023hnt,Saul:2019qnp,UCNA:2019dlk}. 
Even though a combined analysis suggests a 2.86$\sigma$ deviation 
from the SM for Fierz interference \cite{Beck:2023hnt}, 
we adopt the conservative assumption that 
only vector and axial-vector quark operators 
induce non-vanishing nucleon matrix elements.

According to the SM neutrino-induced beta decay, 
the matrix elements of the vector and axial-vector operators 
at the quark-level can be parameterized at the nucleon-level 
as follows \cite{Formaggio:2012cpf} 
(the scalar and tensor components are automatically removed 
because the form factors vanish 
due to the isospin invariance of the strong interaction),
\bea
\label{eq:NME:Vector}
\langle f |  \overline{Q} T^a \gamma^{\mu} Q | i \rangle
&=&
\overline{u}_{N} \bigg[ 
F_{1}(q^2) \gamma^\mu  + 
\frac{i\sigma^{\mu\nu} q_\nu}{2m_{N}} F_{2}(q^2)  
\bigg] u_{N} \,,
\\[2mm]
\label{eq:NME:Axial}
\langle f |  \overline{Q} T^a \gamma^{\mu}\gamma_5 Q  | i \rangle
&=&
\overline{u}_{N} \bigg[ 
F_{A}(q^2) \gamma^\mu\gamma_5 + 
\frac{\gamma_5 q^\mu}{m_{N}} F_{P}(q^2)  
\bigg] u_{N} \,,
\ena
where $M_N = (m_p + m_n)/2$ is the average nucleon mass,  
$q^2 = (p_f - p_i)^2 = (p_e - p_\chi)^2$ is the squared momentum transfer.  
In general, the form factors $F_i(q^2)$ are functions of $q^2$.  
Here we consider only the case where the momentum transfer is negligible 
(\ie, $q^2 \to 0$), and the form factors reduce to the following constants
\cite{Formaggio:2012cpf},
\bea
&&
F_1(0) = 1\,,\qquad\qquad\qquad\qquad\; F_2(0) \simeq 3.706\,,
\\[2mm]
&&
F_A(0)  = -1.2694 \pm 0.0028\,,\quad F_P(0) = \frac{ 2m_N^2 }{m_\pi^2}F_A(0)\,,
\ena
where $m_\pi$ is the mass of the $\pi$-meson. 
Using the matrix elements given in \eqref{eq:NME:Vector} 
and \eqref{eq:NME:Axial}, 
the averaged squared amplitudes at the nucleon level 
can be calculated straightforwardly. 

The full results are relatively complex 
and are given in Appendix \ref{sec:App:Amp}. However, 
the amplitudes in the limit of vanishing threshold mass 
($m_{\text{th}}^{\beta^-} \to 0$) and negligible electron mass ($m_e \to 0$), 
where nucleon masses are approximated as $m_p \approx m_n = m_N$ 
and the electron energy satisfies $E_e \approx m_\chi$, 
can reveal basic properties of the transition probability and are given as follows,
\bea
\overline{|\mathcal{M}_{V}|^{2}}_{ m_{th}^{\beta^-} \to 0 }
&\approx&
\frac{m_N^4}{\varLambda_{V}^{4}} \left[ 
4F_1^2 \frac{ (m_N + m_\chi) m_\chi^2 }{ m_N^3 } +
2( F_1 F_2 + F_2^2 ) \frac{ m_\chi^3 }{ m_N^3 } 
\right] \,,
\\[3mm]
\overline{|\mathcal{M}_{A}|^{2}}_{ m_{th}^{\beta^-} \to 0 }
&\approx&
\frac{m_N^4}{\varLambda_{A}^{4}} \left[ 
4 F_A^2 \frac{ (3m_N + m_\chi) m_\chi^2 }{ m_N^3 } -
4 F_P^2 \frac{ m_\chi^2(m_N + m_\chi) (m_\chi - E_e) }{ m_N^5 } 
\right]\,.~~~~~~~~
\ena
One can clearly see that in this limit the terms proportional to $F_1^2$ ($F_A^2$) 
are suppressed by a factor of $m_\chi^2/m_N^2$ for the (axial-)vector operator. 
Furthermore, for the vector operator, 
the terms proportional to $F_1 F_2$ and $F_2^2$ are suppressed 
by a factor of $m_\chi^3/m_N^3$. 
Hence, the dominant contribution for the vector operator arises 
from the nucleon-level vector current.  
On the other hand, for the axial-vector operator, 
the contribution proportional to $F_P^2$ approaches zero 
in the limit $E_e \to m_\chi$, and the interference term 
(proportional to $F_A F_P$) vanishes exactly. 
Thus, the transition rate is essentially dominated 
by the nucleon-level axial-vector current.

Properties of the amplitude can affect the transition rate due to selection rules. 
Due to angular momentum conservation, 
the daughter nucleus is generally produced in an excited state. 
For small momentum transfers, 
transition rates are conventionally divided into two categories: 
Fermi transitions (nuclear spin unchanged, $\Delta I_F = 0$) and 
Gamow-Teller transitions (nuclear spin changes by $\Delta I_{GT} = \pm1$ 
due to leptonic charged current contributions). 
In general, contributions to these transitions 
are non-trivial linear combinations of terms associated 
with different form factors \cite{Kuramoto:1989tk}. 
However, in the limit $q^2 \to 0$, the connection simplifies: 
Fermi and Gamow-Teller amplitudes relate directly to vector and 
axial-vector currents, respectively. 
Here we study projected sensitivities of experiments listed in 
Table \ref{tab:Exposure}, which offer better sensitivity 
due to larger exposure \cite{Dror:2019dib}.
\begin{table}[th]
\renewcommand\arraystretch{1.58}
\begin{center}
\begin{tabular}{c c c c}
~~~Experiment~~~   & ~~~~~~~~~Taget~~~~~~~~~ & ~~~~~Exposure~~~~~ 
\\\hline\hline
Super-Kamiokande \cite{Super-Kamiokande:2019hga}
&
$\mathrm{H}_2 \mathrm{O}$
& 
$1.71\times 10^5$ t yr
\\[2mm]\hline
\begin{tabular}{l} 
XENONnT \cite{XENON:2023cxc} 
\end{tabular}
& 
Liquid Xe
& 
1.09 t yr 
\\[2mm]\hline
\begin{tabular}{l} 
PandaX-4T \cite{PandaX:2023xgl} 
\end{tabular}
& 
Liquid Xe
& 
 0.55 t yr
\\[2mm]\hline
Borexino \cite{Borexino:2018pev}
& 
$\mathrm{C}_6 \mathrm{H}_3\left(\mathrm{CH}_3\right)_3$
& 
817 t yr
%
%
\\[2mm]
\hline\hline
\end{tabular}
\caption{\it 
Experiments considered here to study projected sensitivities by using 
the induced $\beta^-$ decay.
}
\label{tab:Exposure}
\end{center}
\end{table}
The corresponding lowest thresholds (have the maximum sensitivities) 
of the selected isotopes are given in the Table \ref{tab:Threshold}. 
\begin{table}[th]
\renewcommand\arraystretch{1.58}
\begin{center}
\begin{tabular}{c c c c}
{\rm Process}   & Isotope (Abundance) & ~~~~~$\Delta I = 0$~~~~~ & ~~~~~$\Delta I = \pm1$~~~~~
\\\hline\hline
$ {}^A_{\,6}\mathrm{C} \rightarrow {}^A_{\,7}\mathrm{N} $
&
\begin{tabular}{l} 
$ {}^{12}_{\;\;6}\mathrm{C} $ (98.89\%) \\ $ {}^{13}_{\;\;6}\mathrm{C} $ (1.11\%)
\end{tabular}
& 
\begin{tabular}{r} 18.3\mev \\  2.22\mev \end{tabular}
& 
\begin{tabular}{r} 17.3\mev \\  \;\,5.7\mev \end{tabular}
\\[2mm]\hline
$ {}^A_{\,8}\mathrm{O} \rightarrow {}^A_{\,9}\mathrm{F} $
& 
\begin{tabular}{l} 
$ {}^{16}_{\;\;8}\mathrm{O} $ (99.756\%) \\ 
$ {}^{17}_{\;\;8}\mathrm{O} $ (0.039\%) \\ 
$ {}^{18}_{\;\;8}\mathrm{O} $ (0.205\%)
\end{tabular}
& 
\begin{tabular}{r} 16.4\mev \\  2.76\mev \\  2.70\mev \end{tabular}
& 
\begin{tabular}{r} 16.4\mev \\  3.75\mev \\  1.65\mev \end{tabular}
\\[2mm]\hline
$ {}^{\,A}_{54}\mathrm{Xe} \rightarrow {}^{\,A}_{55}\mathrm{Cs} $
& 
\begin{tabular}{l} 
$ {}^{126}_{\;\;54}\mathrm{Xe} $ (28.4\%) \\ 
$ {}^{131}_{\;\;54}\mathrm{Xe} $ (21.2\%)  \\ 
$ {}^{134}_{\;\;54}\mathrm{Xe} $ (10.4\%)  \\
$ {}^{136}_{\;\;54}\mathrm{Xe} $ (8.8\%) 
\end{tabular}
& 
\begin{tabular}{r} 1.19\mev \\  0.57\mev \\  2.23\mev  \\  1.09\mev \end{tabular}
& 
\begin{tabular}{r} 1.33\mev \\  0.355\mev \\  0.49\mev  \\  1.06\mev \end{tabular}
%
%
\\[2mm]
\hline\hline
\end{tabular}
\caption{\it 
The lowest thresholds of the induced $\beta^-$ decays.
}
\label{tab:Threshold}
\end{center}
\end{table}

\begin{figure}[th]
\centering
\includegraphics[width=0.495\textwidth]{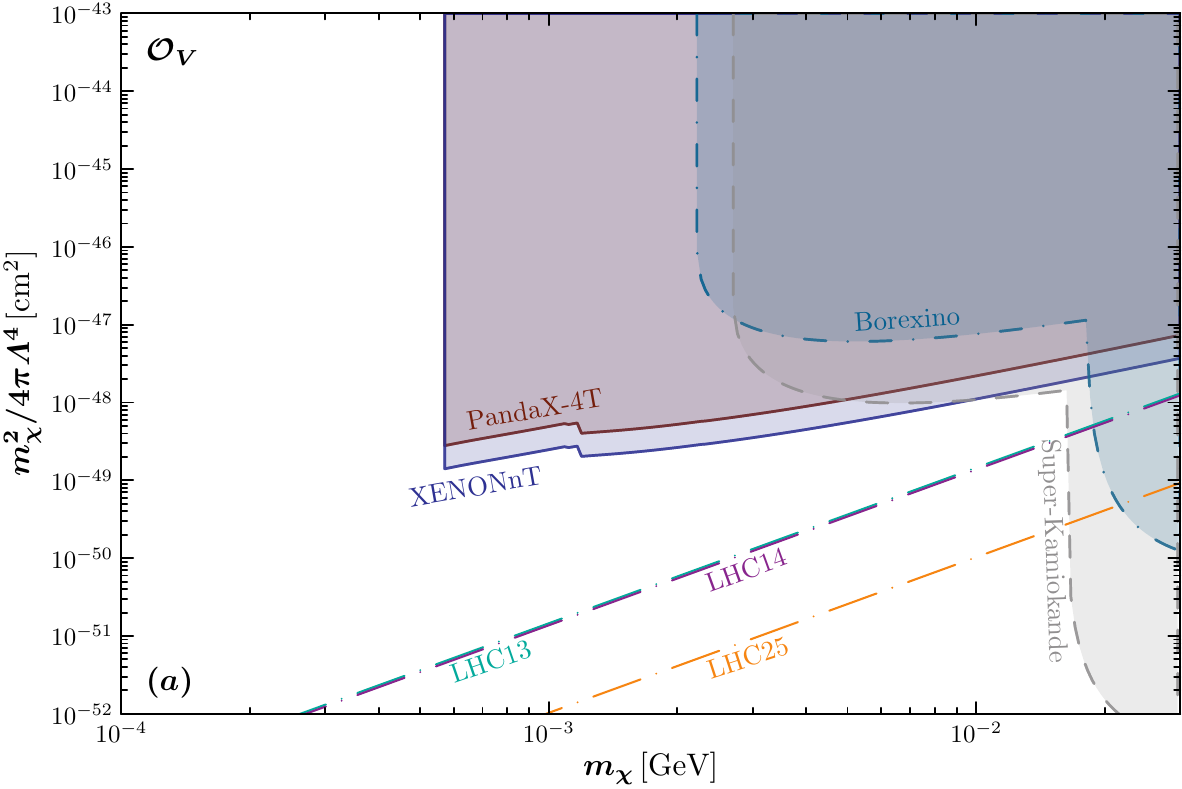}
\hfill
\includegraphics[width=0.495\textwidth]{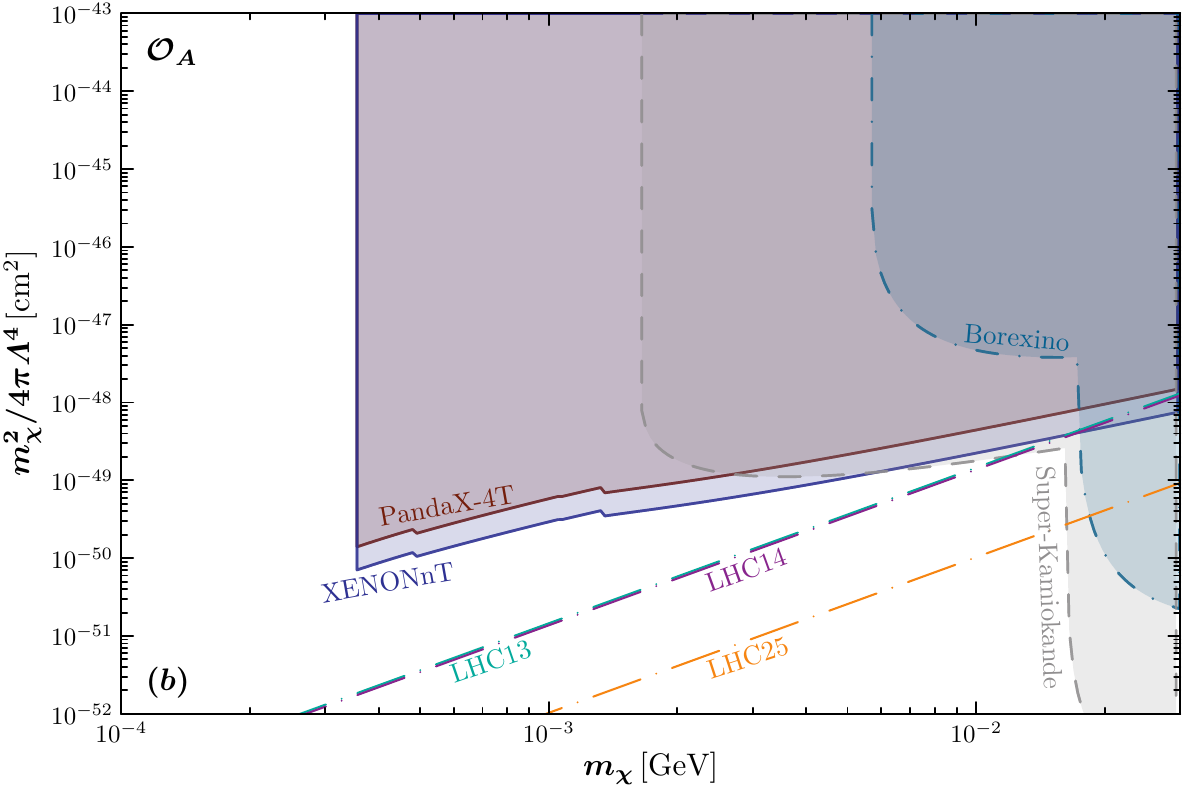}
\caption{\it 
The expected constraints from a dedicated search for induced $\beta^-$ transitions 
at Super-Kamiokande (gray), XENONnT (blue), PandaX-4T (brown), 
and Borexino (midnight blue). The expected exclusion limits at 95\% C.L. at the LHC13 (cyan),
LHC14 (fuchsia) and LHC25 (orange) are also shown. The constraints at the LHC
are obtained by combining all the three processes: mono-$e^\pm$ production, 
and associated productions of the $e^\pm + \gamma/j$ with missing energy.
The \textbf{left} and \textbf{right panel} strand for the vector and axial-vector
operators, respectively.
}
\label{fig:C:IBD}
\end{figure}

Sensitivities of the current experiments are estimated 
by requiring at least 10 events, 
and the results are shown in Fig.~\ref{fig:C:IBD} (a) and (b) 
for the vector and axial-vector operators, respectively. 
For comparison, the expected exclusion limits at 95\% C.L. 
for LHC13 (cyan), LHC14 (fuchsia), and LHC25 (orange) are also shown. 
The constraints at the LHC are obtained by combining all three processes: 
mono-$e^\pm$ production and associated productions of $e^\pm + \gamma$ 
or $j$ with missing energy.  

One can see that direct detection experiments probe different parameter spaces 
due to the use of different isotopes. 
In our range of interest ($m_\chi < 10\,\text{MeV}$), 
the XENONnT experiment gives the strongest constraints. 
However, these regions are already covered by current LHC13 data\footnote{
In contrast, $\sqrt{s}=8\,\text{TeV}$ data gives a weaker bound, $\varLambda \geq 4.5\,\text{TeV}$, as estimated in Ref.~\cite{Dror:2019dib} using data from Ref.~\cite{CMS:2014fjm}.}. 
Hence, hadron colliders are promising for searching signals of four-fermion 
contact interactions involving a dark fermion and a charged lepton.  
The Super-Kamiokande experiment becomes more sensitive 
when $m_\chi \geq 16.4\,\text{MeV}$ 
(the threshold for induced $\beta^-$ decay of ${}_{8}^{16}\mathrm{O}$). 
However, this exceeds the typical nuclear binding energy, 
enabling additional correlated signals that do not rely on nuclear recoils \cite{Dror:2019dib,Lasserre:2016eot} --- for instance, 
emissions of energetic leptons and photons, 
or subsequent $\beta$ decays of the daughter nucleus. 
While we do not explore these signals in detail, 
including them could yield bounds comparable to collider searches.

\subsection{Induced $\beta^+$ Decay}
The nucleus can also undergo dark-fermion-induced $\beta^+$ decay \cite{Vogel:1999zy}. 
Due to Pauli blocking effects in $\beta^+$ decays of heavy isotopes, 
the transition probability to the ground state or lowest-lying excited states 
of the daughter nucleus is heavily suppressed. 
Furthermore, there is no detailed analysis of favored transitions in the literature, 
except for dark-fermion-induced $\beta^+$ decays in hydrogen \cite{Dror:2019dib}. 
Hence, we consider only dark-fermion-induced $\beta^+$ decays in hydrogen
\cite{Vogel:1999zy},
\bee
\overline\chi + {}^{1}_{1}\mathrm{H} \;\to\; n + e^+ \,,
\ene
which has a threshold of $1.8\,\text{MeV}$. 
Here we consider only the Super-Kamiokande and Borexino experiments, 
which contain the hydrogen exposures given in Table \ref{tab:Exposure}. 
The Super-Kamiokande experiment has an energy detection threshold of 
$70\,\text{keV}$; hence, it can probe the dark fermion down to 
a kinematic threshold of $1.87\,\text{MeV}$. 
The Borexino experiment has an energy detection threshold of $3.5\,\text{MeV}$; 
hence, it can probe the dark fermion down to a threshold of $5.3\,\text{MeV}$.

\begin{figure}[th]
\centering
\includegraphics[width=0.495\textwidth]{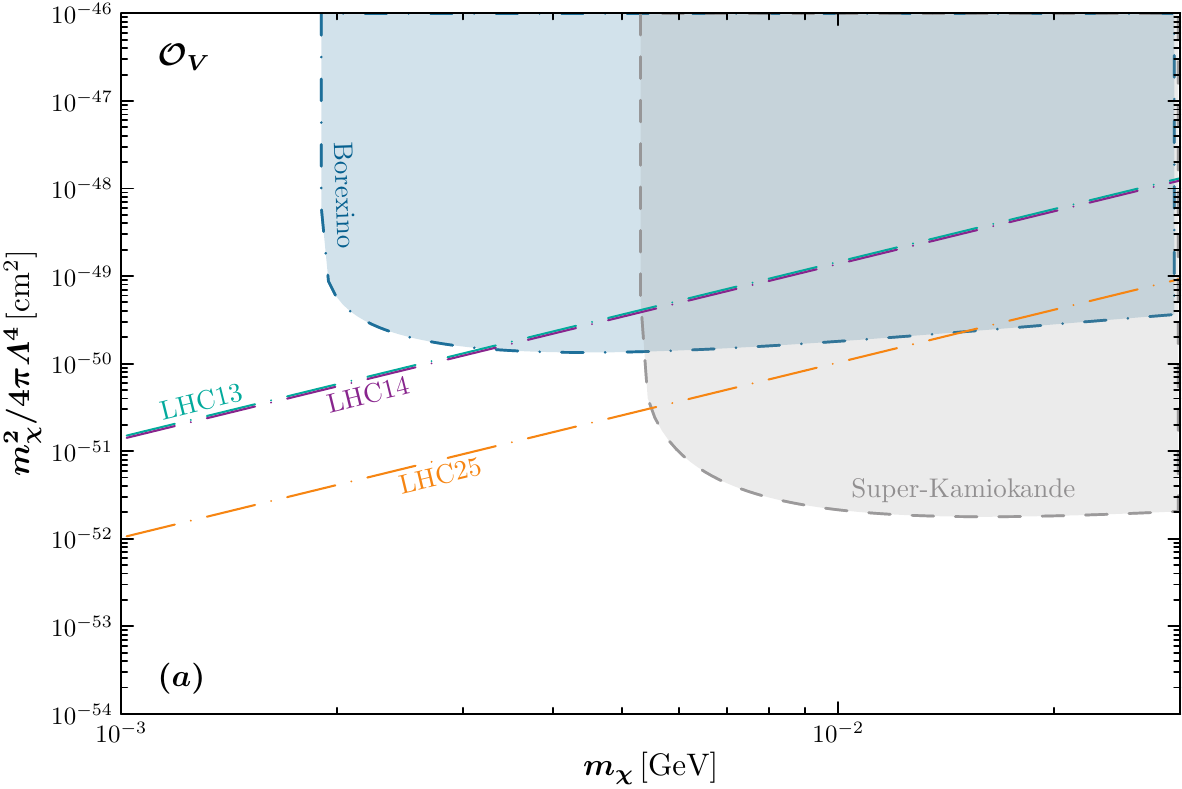}
\hfill
\includegraphics[width=0.495\textwidth]{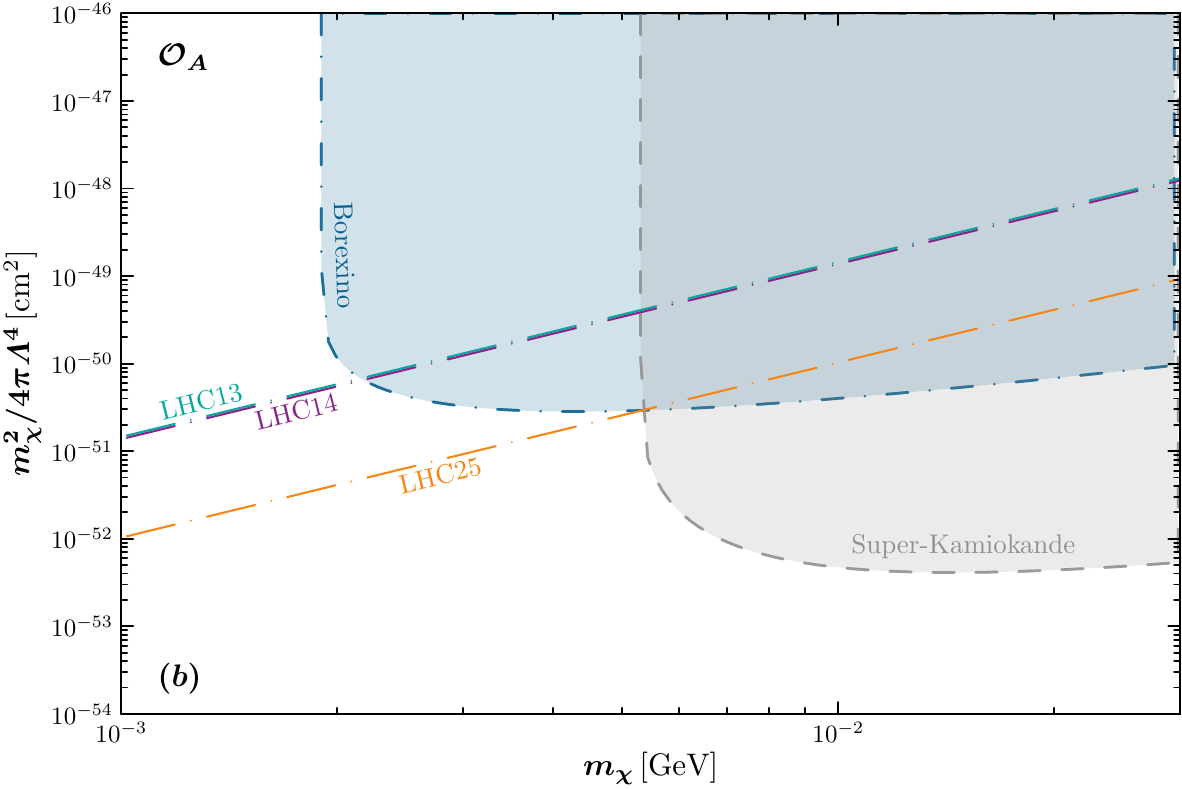}
\caption{\it 
The expected constraints from a dedicated search for induced $\beta^+$ 
transitions at Super-Kamiokande (gray) and Borexino (midnight blue). 
The expected exclusion limits at 95\% C.L. at the LHC13 (cyan),
LHC14 (fuchsia) and LHC25 (orange) are also shown. The constraints at the LHC
are obtained by combining all the three processes: mono-$e^\pm$ production, 
and associated productions of the $e^\pm + \gamma/j$ with missing energy.
The \textbf{left} and \textbf{right panel} strand for the vector and axial-vector
operators, respectively.
}
\label{fig:C:IBPD}
\end{figure}

The matrix element and decay rate can be derived 
from the induced $\beta^-$ decay by replacing $n \leftrightarrow p$. 
The expected exclusion limits are obtained by requiring at least 10 events, 
and the results are shown in Fig.~\ref{fig:C:IBPD} (a) and (b) for the vector 
and axial-vector operators, respectively. 
Since induced $\beta^+$ decays have relatively large thresholds, 
they provide complementarity to induced $\beta^-$ decay searches.  
On the other hand, LHC constraints are weaker in parts of the parameter space 
than bounds from induced $\beta^+$ decays --- 
this contrasts with $\beta^-$ decays, 
where LHC constraints dominate almost the entire parameter space. 
In this sense, collider searches and induced $\beta^\pm$ decays 
are complementary in probing the dark fermion.

\section{Conclusion}
\label{sec:conclusion}
For a fermionic dark matter particle, 
the effective interaction with SM particles at leading order 
can be simply described by four-fermion contact operators. 
Here we focus on operators involving a dark matter fermion, 
a charged lepton, and quark pairs, 
which are explicitly defined in Eq.~\eqref{eq:effo}. 
We study signatures of these operators at the LHC 
and in direct detection experiments.

At the LHC, we show that the mono-electron production is a powerful probe 
of the four-fermion contact couplings. 
This is simply because the irreducible background 
dominantly arises from resonant decays of the charged gauge bosons $W^\pm$, 
and thus backgrounds can be significantly reduced 
by applying a kinematic cut on the transverse mass $m_T$.  
We also studied associated productions of 
an electron and a photon/jet with missing energy. 
Since the total cross sections are suppressed by factors of $\alpha_{\rm EM}$ 
and $\alpha_s$, respectively, the corresponding exclusion limits are weaker 
than those from mono-$e^\pm$ production. 
However, these processes provide more observables, and 
correlations between them may help distinguish the structures of signal operators. 
Furthermore, combining all three channels enhances the exclusion limits. 
For (axial)-vector operators, using current LHC data at $\sqrt{s}=13\tev$, 
the combined lower bounds on the energy scales are 
$\Lambda_{V/A} \geq 12.1\tev$ for a massless dark fermion.  
Given that signal cross sections grow with the center-of-mass energy 
$\sqrt{\hat{s}}$, we also study future sensitivities for the LHC 
at $\sqrt{s}=14\tev$ and $25\tev$.

For direct detection experiments, 
by matching operators between the quark and hadron levels, 
we show that vector and axial-vector currents come into play separately 
(this also depends on the parameterization of the four-fermion contact operator). 
We focus on specific induced beta decays to probe signals: 
vector currents mediate Fermi transitions, 
while axial-vector operators induce Gamow-Teller transitions.  
For induced $\beta^-$ decay, 
we study expected constraints from Super-Kamiokande, XENONnT, PandaX-4T, 
and Borexino experiments, which exhibit better sensitivities 
in distinct parameter spaces. For induced $\beta^+$ decay, 
we consider only Super-Kamiokande and Borexino 
for the transition $\overline{\chi} + {}_{1}^{1}\mathrm{H} \to n + e^+$. 
We find that induced $\beta^+$ decay provides better sensitivities 
when the dark fermion mass is relatively large, 
e.g., $m_\chi > 1.8\,\text{MeV}$ (the threshold for the above transition).

By comparing constraints from the LHC, 
we have shown that collider searches can provide better sensitivities 
than those from induced $\beta^-$ decay 
across almost the entire parameter space, as shown in Fig.~\ref{fig:C:IBD}. 
In contrast, induced $\beta^+$ decay is more sensitive than collider searches 
for a relatively heavy dark fermion, as shown in Fig.~\ref{fig:C:IBPD}.  
In our analysis of induced beta decay, 
we assumed $m_\chi < 10\,\text{MeV}$ (the typical nuclear binding energy). 
Beyond this value, additional correlated signals that do not rely on nuclear recoils 
can appear \cite{Dror:2019dib,Lasserre:2016eot} --- for instance, 
emissions of energetic leptons (electrons/positrons) and photons, 
or recoil and subsequent $\beta$ decay of the daughter nucleus. 
While we did not study these signals in detail, 
we stress that including them could establish complementarity across 
the entire parameter space between collider searches and direct detections.

\appendix
\section{Nucleon Level Squared Amplitudes}
\label{sec:App:Amp}
The averaged squared amplitudes at the nucleon level are given as,
\bea
&& \nonumber
\overline{|\calm_V|}
\\[3mm] 
&=& \nonumber
\frac{ m_n^4 }{\varLambda_V^4}  \bigg\{
F_1^2 \bigg[ \bigg( \frac{8m_\chi^2}{m_n^3} + \frac{4m_\chi (2m_n - m_p) }{m_n^3} \bigg) p_{e} -  \frac{4m_\chi  }{m_n^3}p_e^2  \bigg]
\\[3mm]
&+& \nonumber
F_2^2 \bigg[ \frac{4m_\chi^4 + 2m_\chi^2 (3m_n + m_p)^2 + 2 m_\chi(m_n^2 -m_p^2)(5m_n + 3m_p) + 12m_\chi^3 m_n }{m_n^3 (m_n + m_p)^2} p_e ~~~~~~~~~~
\\[3mm]
&-& \nonumber
 \frac{8m_\chi^3 m_n + 16m_\chi^2 m_n^2  + m_\chi(14 m_n^3 + 8m_n^2 m_p+ 2 m_n m_p^2) }{m_n^4 (m_n + m_p)^2} p_e^2
+
\frac{4 m_\chi^2 + 4 m_\chi m_n }{m_n^3 (m_n + m_p)^2} p_e^3
\bigg]~~~~
\\[3mm]
&-&
\frac{2F_1F_2 m_\chi m_n p_e  }{ m_n^4 (m_n + m_p) } \bigg[  2 m_\chi^2 - m_n^2 
+ m_p^2 + m_\chi(m_n- 9 m_p - 2p_e) - m_n p_e + 7 m_p p_e \bigg] \bigg\} \,,
\ena
\bea
&& \nonumber
\overline{|\calm_A|}
\\[3mm] 
&=& \nonumber
\frac{ m_n^4 }{\varLambda_A^4}  \bigg\{
F_A^2 \bigg[
\frac{ 8 m_\chi^2 + 4 m_\chi ( 2m_n + m_p) }{ m_n^3 } p_e - \frac{4m_\chi}{m_n^3} p_e^2
\bigg]
\\[3mm]
&-& \nonumber
F_P^2 \bigg[ \bigg( \frac{ 16 m_\chi^4 + 16 m_\chi^3 (2m_n -m_p) + 8 m_\chi^2(3m_n^2
-2m_nm_p - m_p^2)  }{ m_n^3 (m_n + m_p)^2 }
\\[3mm]
&+& \nonumber
\frac{  8m_\chi(m_n^3 -m_n^2m_p-m_n m_p^2 + m_p^3) }{ m_n^3 (m_n + m_p)^2 }
\bigg) p_e
+
\bigg( \frac{-32 m_\chi^3 - 16 m_\chi^2(3m_n -m_p) }{ m_n^3 (m_n + m_p)^2 }
\\[3mm]
&+& 
 \frac{ 8 m_\chi ( -3m_n^2 + 2m_nm_p + m_p^2  }{ m_n^3 (m_n + m_p)^2 } \bigg) p_e^2
 +
 \frac{16 m_\chi (m_\chi  + m_p)}{ m_n^3 (m_n + m_p)^2 } p_e^3 \bigg] \bigg\} \,.
\ena

\acknowledgments
K. M. was supported by the Natural Science Basic Research Program of Shaanxi (Program No. 2023-JC-YB-041), and the Innovation Capability Support Program of Shaanxi (Program No. 2021KJXX-47).

\bibliographystyle{mDraft}
\bibliography{MonolR2}

\end{document}